\documentclass{article}

\usepackage[T1]{fontenc}
\usepackage[utf8]{inputenc}
\usepackage{hyperref}
\usepackage{spconf,amsmath,graphicx}
\usepackage{float}
\usepackage{amsmath}
\usepackage{bm}
\usepackage{mathtools}
\usepackage[font=normalsize,skip=12pt]{caption,subcaption}
\usepackage[backend=biber,style=ieee,natbib=false,mincrossrefs=1000]{biblatex}
\usepackage{fancyhdr}
\usepackage{xcolor}
\usepackage{xspace}
\usepackage{makecell}
\usepackage{booktabs}
\usepackage{tabulary}
\usepackage[title]{appendix}
\usepackage{calc}
\usepackage{tikz}
\usetikzlibrary{spy,calc,positioning}
\usepackage{adjustbox}
\usepackage{bold-extra}
\usepackage[pdftex,outline]{contour}

\contourlength{0.6pt}
\newcolumntype{P}[1]{>{\centering\arraybackslash}p{#1}}

\fancypagestyle{first_page_style}
{
   \fancyhf{}
   \lfoot{\color{gray}\scriptsize Copyright 2020 IEEE. Published in the IEEE 2020 International Conference on Image Processing (ICIP 2020), scheduled for 25-28 October 2020 in Abu Dhabi, United Arab Emirates. Personal use of this material is permitted. However, permission to reprint/republish this material for advertising or promotional purposes or for creating new collective works for resale or redistribution to servers or lists, or to reuse any copyrighted component of this work in other works, must be obtained from the IEEE. Contact: Manager, Copyrights and Permissions / IEEE Service Center / 445 Hoes Lane / P.O. Box 1331 / Piscataway, NJ 08855-1331, USA. Telephone: + Intl. 908-562-3966.}
}

\newcommand{\shrinksection}{\vspace{-1.5mm}}
\newcommand{\shrinkcaption}{\vspace{-4.6mm}}

\newlength\figwidth
\newlength\imagewidth
\newlength\subcap

\newcommand{\tines}{\!\,\times\!\,}

\title{Channel-wise Autoregressive Entropy Models\\for Learned Image Compression}

\name{David Minnen \& Saurabh Singh}
\address{Google Research, Mountain View, CA 94043, USA}

\begin{document}
\maketitle

\begin{abstract}
In learning-based approaches to image compression, codecs are developed by optimizing a computational model to minimize a rate-distortion objective. Currently, the most effective learned image codecs take the form of an entropy-constrained autoencoder with an entropy model that uses both forward and backward adaptation. Forward adaptation makes use of side information and can be efficiently integrated into a deep neural network. In contrast, backward adaptation typically makes predictions based on the causal context of each symbol, which requires serial processing that prevents efficient GPU / TPU utilization. We introduce two enhancements, channel-conditioning and latent residual prediction, that lead to network architectures with better rate-distortion performance than existing context-adaptive models while minimizing serial processing. Empirically, we see an average rate savings of 6.7\% on the Kodak image set and 11.4\% on the Tecnick image set compared to a context-adaptive baseline model. At low bit rates, where the improvements are most effective, our model saves up to 18\% over the baseline and outperforms hand-engineered codecs like BPG by up to 25\%.


\end{abstract}


\fancyhf{}
\renewcommand{\headrulewidth}{0pt}
\setcounter{page}{1}
\cfoot{\thepage}
\thispagestyle{first_page_style}

\makeatletter
\DeclareRobustCommand\onedot{\futurelet\@let@token\@onedot}
\def\@onedot{\ifx\@let@token.\else.\null\fi\xspace}

\def\eg{\emph{e.g}\onedot} \def\Eg{\emph{E.g}\onedot}
\def\ie{\emph{i.e}\onedot} \def\Ie{\emph{I.e}\onedot}
\def\cf{\emph{c.f}\onedot} \def\Cf{\emph{C.f}\onedot}
\def\etc{\emph{etc}\onedot} \def\vs{\emph{vs}\onedot}
\def\wrt{w.r.t\onedot} \def\dof{d.o.f\onedot}
\def\etal{\emph{et al}\onedot}
\makeatother

\section{Introduction}
\label{sec:intro}
\shrinksection


Most recent research in learned image compression uses deep neural networks, and a wide range of model architectures have been explored including recurrent networks~\cite{toderici2017cvpr, baig2017, minnen2017icip, johnston2018cvpr} and autoencoders with an entropy-constrained bottleneck~\cite{theis2017iclr, balle2017iclr, rippel2017icml, li2017importance, minnen2018icip, balle2018iclr, mentzer2018cvpr, klopp2018bmvc, minnen2018neurips, lee2019cae, zhou2019clic, wen2019clic}. In models that use an autoencoder, an \textit{analysis} network transforms pixels into a quantized latent representation suitable for compression by standard entropy coding algorithms, while a \textit{synthesis} network is jointly optimized to transform the latent representation back into pixels. 

\begin{figure*}[t]
  \centering
  \includegraphics[width=\linewidth]{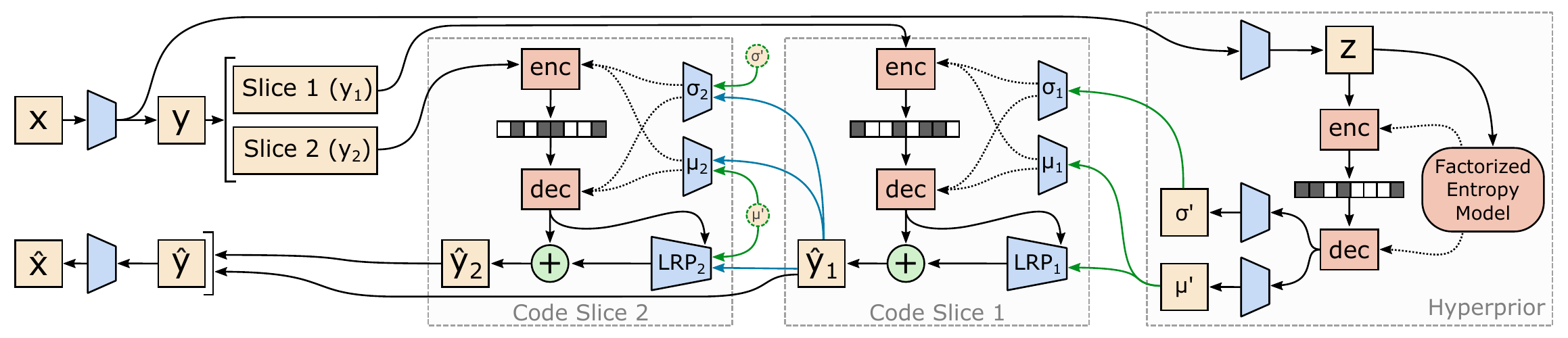}
  \shrinkcaption
  \vspace{-3mm}  
  \caption{This data-flow diagram shows the architecture of our compression model with latent residual prediction (LRP) and two slices for channel-conditioning (CC). Tan blocks represent data tensors, blue represents transforms composed of convolutional layers, green is for basic arithmetic operations, and red represents entropy coding. In this model, an input image ($x$) is transformed into a latent representation ($y$) before being split along the channel dimension. The first slice ($y_1$) is compressed using a Gaussian entropy model conditioned solely on the hyperprior (green arrows from $\mu'$ and $\sigma'$), while the entropy model for the second slice ($y_2$) is conditioned on both the hyperprior and the decoded symbols in the first slice (blue arrows from $\hat{y}_1$). After each slice is quantized and range coded (\texttt{enc} and \texttt{dec} blocks), quantization error is reduced by adding the predicted residual (\texttt{LRP\textsubscript{1}} and \texttt{LRP\textsubscript{2}}), which is conditioned on the hyperprior via $\mu'$. Finally, the decoded slices ($\hat{y}_1$ and $\hat{y}_2$) are concatenated to form $\hat{y}$ and transformed into the final reconstructed image ($\hat{x}$).}
  \label{fig:arch}
\end{figure*}

To date, the most effective models make use of both \textit{forward} and \textit{backward-adaptive} components to improve the predictive power of the entropy model, which leads to higher compression rates without increasing distortion. Forward-adaption typically makes use of side information, for example in the form of local histograms over the quantized latent representation~\cite{minnen2018icip} or a learned hyperprior~\cite{balle2018iclr}. The hyperprior approach is particularly popular since it can easily be integrated into an end-to-end optimized network and allows for efficient encoding and decoding.

Backward-adaptation, on the other hand, typically incorporates predictions from the causal context of each symbol, \ie neighboring symbols above and to the left of the current symbol as well as symbols in previously decoded channels~\cite{klopp2018bmvc, lee2019cae, minnen2018neurips, mentzer2018cvpr}. In such context-adaptive models, encoding can still be performed efficiently using masked convolution, which will run in parallel across the entire latent tensor on a GPU or TPU~\cite{vanDenOord2016pixelcnn}. Decoding, however, is inherently serial, and thus does not effectively utilize massively parallel hardware.



Our goal is to develop an image compression architecture capable of matching the rate-distortion (RD) performance of a context-adaptive model while minimizing serial processing that can lead to slow decoding times. Toward this goal, we explore two architectural enhancements: channel-conditioning (CC) and latent residual prediction (LRP). In addition, we show how training synthesis transforms with rounded latent values interacts positively with CC and LRP to further boost RD performance.

The combined effect of these improvements is a highly parallelizable architecture that outperforms recently proposed context-adaptive models~\cite{minnen2018neurips, klopp2018bmvc, lee2019cae} by 6.7\% on Kodak~\cite{kodak} and 11.4\% on the Tecnick image set~\cite{tecnick}. We see even larger gains compared to standard codecs and learning-based models that do not use context (see Figures~\ref{fig:kodak-rd} and~\ref{fig:kodak-bdbins}). The coding improvements provided by CC and LRP are most effective at low bit rates where our model saves more than 16\% compared to the context-adaptive baseline and as much as 25\% relative to BPG~\cite{bpg}. The following three sections describe channel-conditioning, latent residual prediction, and round-based training. A detailed analysis of the empirical results is presented in Section~\ref{sec:results} and discussed in Section~\ref{sec:discussion}.

\section{Channel-Conditional Entropy Models}
\label{sec:cc}
\shrinksection


Our model builds on the hyperprior architecture introduced in~\cite{balle2018iclr}. This model learns to generate an image-dependent hyper-latent tensor that is compressed and transmitted as side information. It jointly learns to transform this tensor into the entropy parameters used to compress the symbols that represents the input image (see the \textit{Hyperprior} block at the right of Figure~\ref{fig:arch}). Hyperprior models typically use a conditional Gaussian model parameterized by scale~\cite{balle2018iclr} or both scale and mean, and the most effective models combine information from the hyperprior (forward-adaptation) with a spatially autoregressive model (backward-adaptation) before predicting the entropy parameters $\mu$ and $\sigma$~\cite{klopp2018bmvc, minnen2018neurips, lee2019cae}.

Conditioning on the causal context allows for better modeling of spatial correlation and is commonly used in standard image codecs~\cite{webp,bpg,jpeg2000} and for intra-frame prediction in video codecs~\cite{hevc,av1,avc}. In a learning-based codec, the model must estimate the parameters of a spatially autoregressive (AR) model. This approach is effective but requires running the AR model sequentially to decode each symbol, which can slow down decoding times on GPUs and TPUs compared to architectures that better utilize the massively parallel processing abilities of such hardware. For this reason, we explore channel-conditional (CC) models, which split the latent tensor along the channel dimension into $N$ roughly equal-size slices, and conditions the entropy parameters for each slice on previously decoded slices.

Figure~\ref{fig:arch} provides a high-level overview of this architecture where the blue arrows show how $y_2$ (the second slice) is conditioned on $\hat{y}_1$ (the first slice). In a model with more splits, the third slice ($y_3$) would be conditioned on the hyperprior along with both $\hat{y}_1$ and $\hat{y}_2$, \etc.

We can interpret CC models as autoregressive along the channel dimension rather than the spatial dimensions. Although this structure also introduces some serial processing (slice $y_i$ can only be decoded after slices $[y_1 \ldots y_{i-1}]$), we typically use relatively few slices due to diminishing benefits to RD performance (see Figure~\ref{fig:num-splits}). Note that in a model with $N$ slices, each slice contains $W \times H \times \frac{C}{N}$ values that can be processed in parallel (where $W$, $H$, and $C$ correspond to the width, height and number of channels, respectively). Contrast with a spatially autoregressive model where a naive implementation requires $W \times H$ sequential steps with only $C$ values computed during each run. A more careful implementation using wavefront processing adds some parallelization~\cite{alvarez2012wavefront} but still far less than channel-conditioning.


\section{Latent Residual Prediction}
\label{sec:lrp}
\shrinksection

Autoencoder models learn to transform pixel values ($x$) into real-valued latents ($y$) that are quantized before they are losslessly compressed. This process inevitably leads to a residual error in the latent space ($r = y - Q[y]$) that manifests as extra distortion when $Q[y]$ is transformed back into the pixel domain ($\hat x$).

Latent residual prediction attempts to reduce this quantization error by predicting the residual based on the hyperprior and any previously decoded slices. The predicted residual is added to the quantized latents slice-by-slice, which allows LRP to improve results both by decreasing distortion and by decreasing entropy since the entropy parameters used to code later slices are conditioned on previous ones that include LRP.

Previous approaches for augmenting the input to the synthesis transform either re-used the mean prediction directly~\cite{zhou2019clic} or used dilated convolution to provide additional features based on a larger receptive field~\cite{wen2019clic}. In both cases, however, the extra features were concatenated with the latent tensor, which increases computation, and neither used channel-conditioning, which means that potential improvements could only affect distortion.

\section{Training with Rounded Latent Values}
\label{sec:train-round}
\shrinksection

All compression models trained using gradient-based optimization are hindered by quantization, which yields gradients that are either zero or infinite at all values. Typically, researchers avoid this problem by either training with uniform noise, which simulates ``noisy quantization'' without destroying the gradient~\cite{zamir2014, BaLaSi16a, balle2017iclr, balle2018iclr, klopp2018bmvc, minnen2018neurips, lee2019cae}, or they use straight-through gradients where rounding is applied but the true gradient function is replaced with the identity function~\cite{theis2017iclr}.

Although space constraints preclude a full report on the effects of different training methods, we empirically found that a mixed approach improves RD performance. Our baseline models replace quantization with uniform noise during training: $Q[y] \doteq y + \mathcal{U}(-\frac{1}{2}, \frac{1}{2})$. The mixed approach uses the same uniform noise for learning entropy models but replaces the noisy tensor with a rounded one whenever the quantized tensor is passed to a synthesis transform. Looking at Figure~\ref{fig:arch}, the difference is essentially whether the quantized tensor is flowing to the right (add noise) or left (round with straight-through gradients). We experimented with using the rounding-based method everywhere, but this approach performed worse than the noise-based baseline.

\section{Experimental Results}
\label{sec:results}
\shrinksection

\begin{figure}[tb]
  \centering
  \includegraphics[width=\columnwidth]{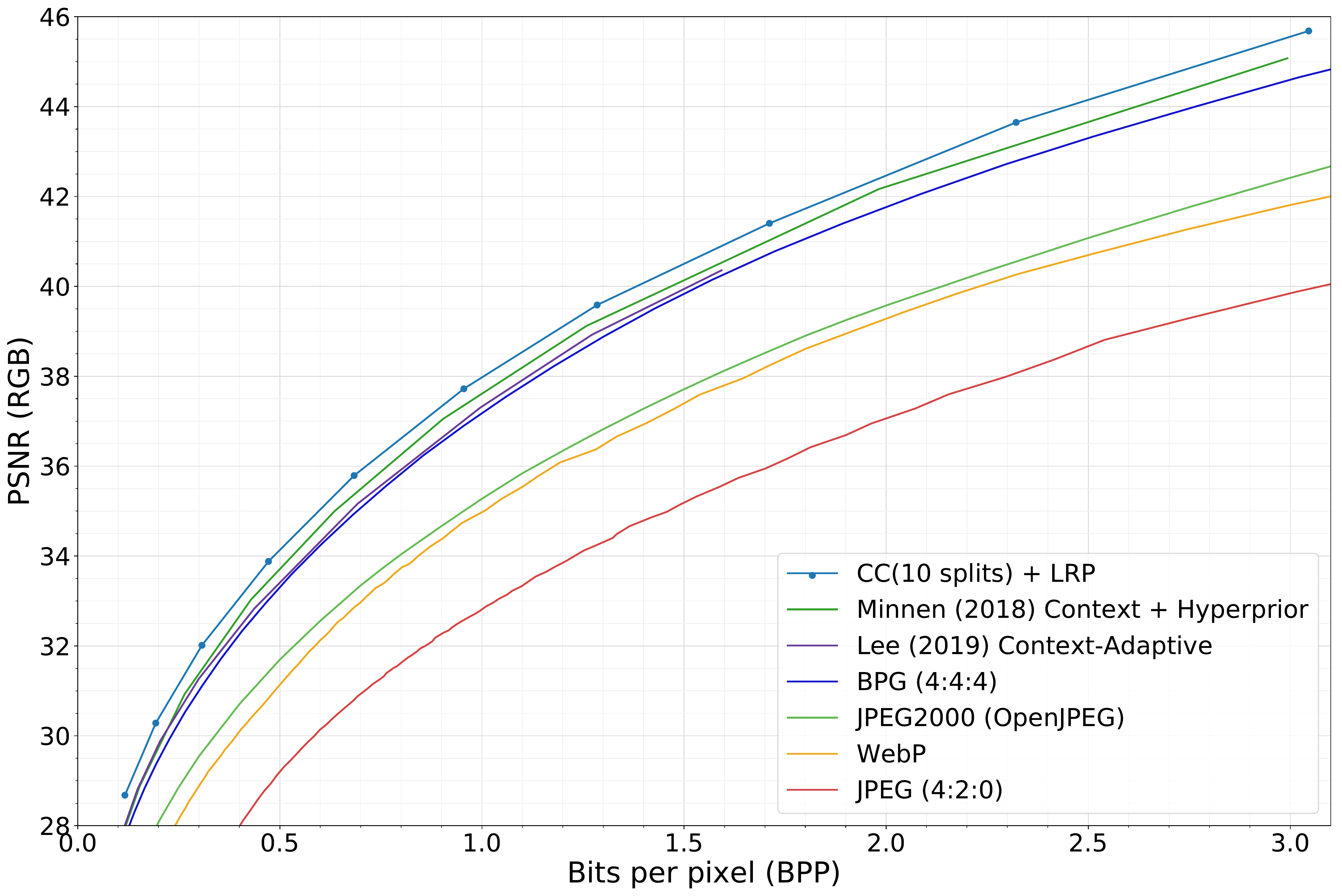}
  \shrinkcaption
  \caption{Models using channel-conditioning and latent residual prediction outperform both the learning-based baselines and standard codecs on the Kodak image set.}
  \label{fig:kodak-rd}
\end{figure}

\begin{figure}[tb]
  \centering
\includegraphics[width=\columnwidth]{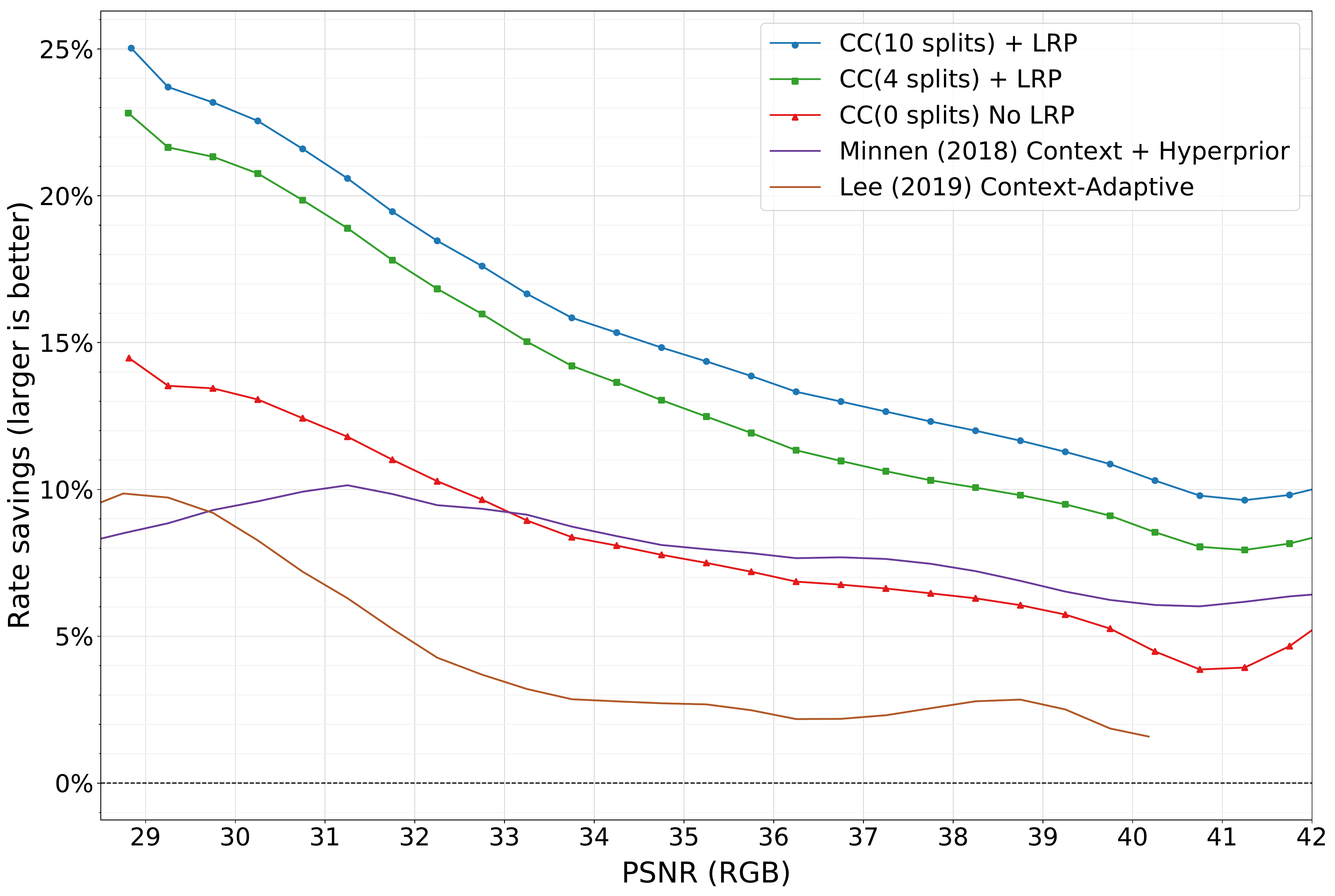}
  \shrinkcaption
  \caption{Each curve shows the rate savings relative to BPG averaged over the Kodak image set. Our largest model (10 CC splits + LRP + round-based training) outperforms BPG by 10\% at high bit rates and up to 25\% at low bit rates.}
  \label{fig:kodak-bdbins}
\end{figure}

In this section, we evaluate the effects of using CC, LRP, and round-based training in a learned image codec. Figure~\ref{fig:kodak-rd} compares RD curves averaged over the Kodak image set~\cite{kodak}. The graph shows that our full model (10 CC slices + LRP + round-based training) outperforms all of the standard codecs (BPG, JPEG2000, WebP, and JPEG) as well as learning-based codecs that combine spatial context with a hyperprior~\cite{minnen2018neurips, lee2019cae}. To improve clarity, earlier learning-based methods, including~\cite{johnston2018cvpr, theis2017iclr, balle2017iclr, balle2018iclr, mentzer2018cvpr, minnen2017icip, baig2017, minnen2018icip, klopp2018bmvc, rippel2017icml, li2017importance}, are not shown in Figure~\ref{fig:kodak-rd}, but all of these methods have worse RD performance than both BPG and our CC + LRP model.


Additional results are shown in Figure~\ref{fig:kodak-bdbins}, which plots the relative rate savings compared to BPG at different quality levels. Larger values correspond to larger relative rate savings and thus better compression. This graph generalizes a Bjøntegaard Delta (BD) chart~\cite{VCEG-M33} by plotting rate savings as a function of quality, rather than only presenting the average savings. Our largest model, which uses 10 CC slices, provides a significant rate savings over BPG, ranging from 10\% at higher quality levels up to 25\% at the lowest. This corresponds to an average BD rate savings of 13.9\% over BPG and 6.7\% over the context-adaptive baseline~\cite{minnen2018neurips}. The following sections analyze how each proposed improvements contributes to the final result.


\subsection{Number of Channel-Conditional Slices}
\label{subsec:splits}

\begin{figure}[tb]
  \centering
  \includegraphics[width=\columnwidth]{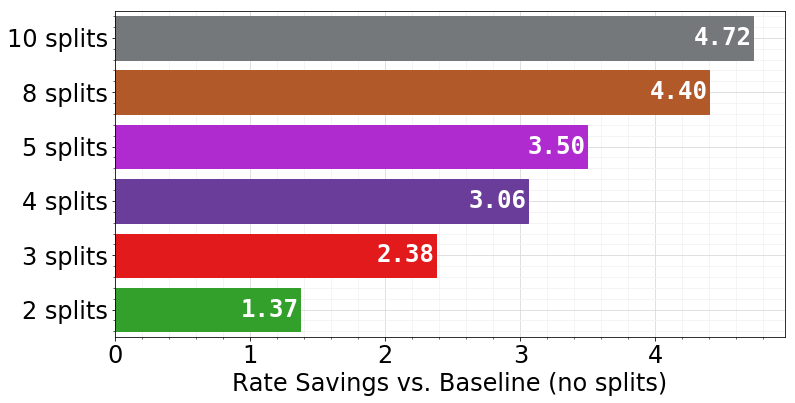}
  \shrinkcaption
  \caption{RD performance increases with additional channel-conditional splits. The graph shows BD rate savings for models that are identical except for the number of CC splits. Note that these models were trained without LRP to isolate the effect of channel-conditioning.}
  \label{fig:num-splits}
\end{figure}

Figure~\ref{fig:num-splits} shows the average rate savings as the number of channel-conditioning slices increases. When we split the latent tensor into more slices, there are more opportunities to model the dependencies between channels, which reduces entropy. This benefit, however, comes at the cost of extra computation, and we also see diminishing returns as the number of slices increases.


\subsection{Latent Residual Prediction}
\label{subsec:lrp}

\begin{figure}[t]
  \centering
  \includegraphics[width=\columnwidth]{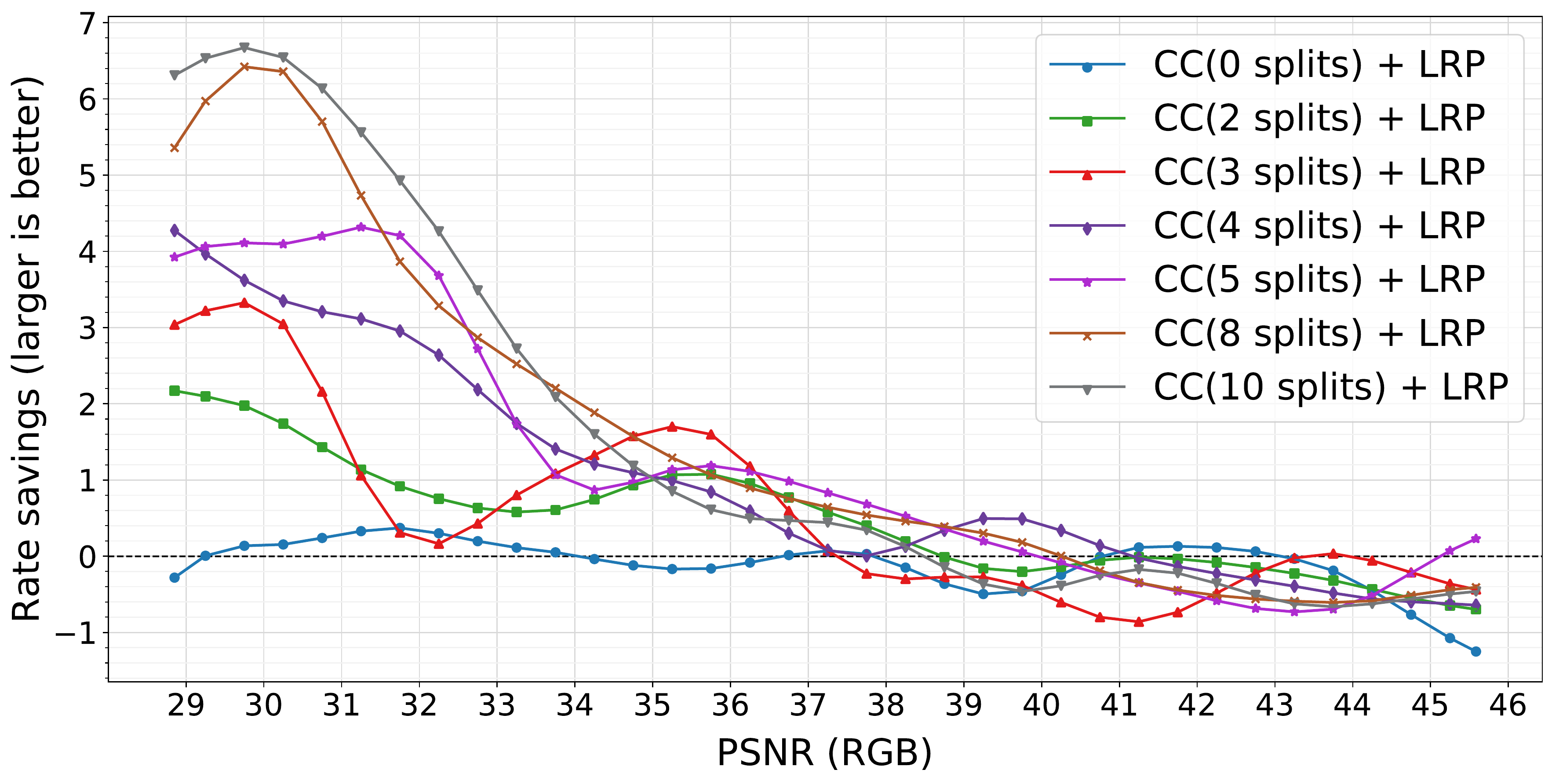}
  \shrinkcaption
  \caption{Combined with channel-conditioning, latent residual prediction (LRP) helps significantly at lower bit rates but reduces performance slightly at the highest bit rates.}
  \label{fig:lrp}
\end{figure}

Figure~\ref{fig:lrp} shows the effect of LRP for different numbers of channel-conditioning splits. Each curve compares a model trained with LRP to an identical model without LRP by plotting the relative rate savings when LRP is used.

The figure shows several effects. First, LRP has almost no benefit for models that do not use channel-conditioning, which we can see because the blue ``CC(0 splits)'' curve is always close to zero. Second, regardless of the number of CC splits, LRP slightly reduces RD performance at high bit rates. At low bit rates, however, the benefit of LRP increases with the number of CC slices and improves compression by more than 6\% for the model with 10 splits.


\subsection{Rounding-based Optimization}
\label{subsec:train-round}

\begin{figure}[b]
  \centering
  \includegraphics[width=\columnwidth]{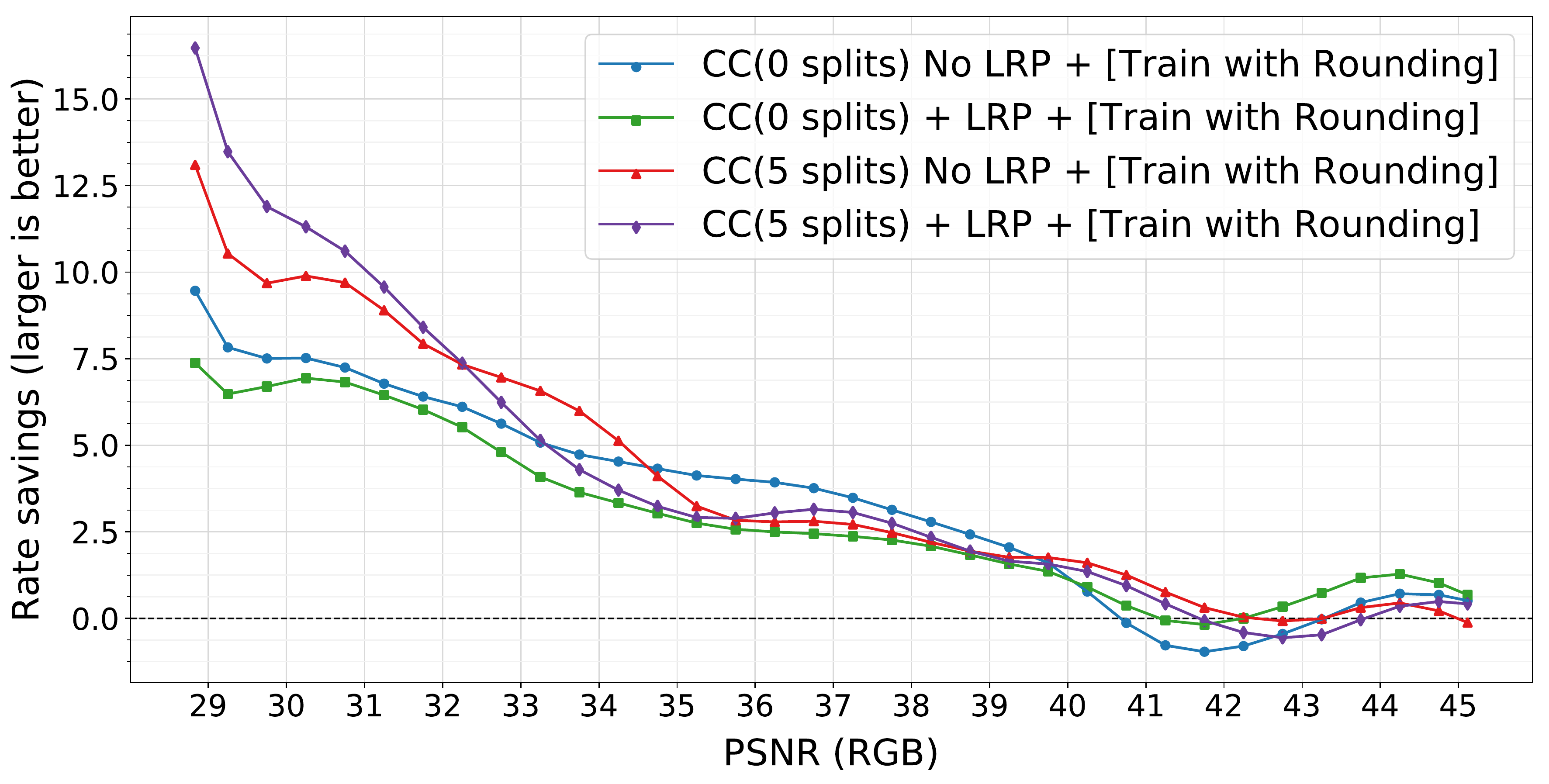}
  \shrinkcaption
  \caption{Each curve shows the average rate savings on the Kodak image set when training part of the model with rounded values vs. using uniform noise everywhere (see Section~\ref{sec:train-round} for details). At low and moderate bit rates, there is a significant benefit to round-based training.}
  \label{fig:noise-vs-round}
\end{figure}

Figure~\ref{fig:noise-vs-round} shows the impact of mixed training with noise and round-based handling of quantized tensors as described in Section~\ref{sec:train-round}. The figure shows results for two CC models (zero and five splits) and plots both variants with and without LRP. Each curve shows the rate savings relative to an identical model optimized using uniform-noise everywhere, which means that the rate savings are due entirely to the change in how quantization is handled. We see the same trend in all cases: the benefit is minimal at higher quality levels but becomes significant at lower bit rates. For the ``CC (5 splits) + LRP'' model, the savings exceed 15\% at the lowest bit rates.


\section{Discussion}
\label{sec:discussion}
\shrinksection

From a theoretical perspective, the positive results from both CC and LRP are somewhat surprising. Ideally, the optimization process should expand the range of each channel to balance the rate-distortion trade-off, which means that using additional bits in the hyperprior to drive LRP would not be helpful. Essentially, channels that significantly reduce distortion would use more symbols, which can be interpreted as finer precision, \eg consider a channel that uses values $[-1, 0, 1]$ vs. one that uses $[-100, -99, \ldots, 99, 100]$ and is scaled by $\frac{1}{100}$ in the next convolutional layer. Since the most useful channels should already have higher effective granularity, there is less opportunity for LRP to provide a benefit.

Similarly, the analysis transform would ideally learn to map pixels into a latent space such that each channel is conditionally independent given the hyperprior. If this is not the case, it means there is redundant information, which will increase entropy without reducing distortion.

Empirically, we see significant improvements using both CC and LRP, which implies that existing models are far from ideal. Further research is needed to understand why the models are failing to reach an optimal state, but we can theorize that the relatively simple 4-layer convolutional networks that make up the analysis and synthesis transforms lack the capacity to generate/decode a latent representation with conditionally independent channels. Alternatively, the networks may have the necessary capacity, but our learning procedure, which uses the Adam optimizer~\cite{KiBa15}, is unable to find a suitable minimum despite training for five million steps.

By combining channel-conditioning, latent residual prediction, and round-based training, we have developed a neural image compression architecture that outperforms a corresponding context-adaptive model while minimizing serial processing. In future research, we plan to investigate combining channel-conditioning with spatial context modeling to see if the two approaches are complementary.

\vfill
\clearpage


\twocolumn[  
    \begin{@twocolumnfalse}
        \begin{center}
             {\Large\bfseries{\scshape{Appendix \& Supplemental Material}}}
        \end{center}
        \vspace{8mm}
    \end{@twocolumnfalse}
]


\appendixtitleon
\begin{appendices}

\section{Architecture Details}

\begin{table*}[tb]
  \centering
  \scriptsize
  \setlength\extrarowheight{1pt}
  \setlength\tabcolsep{3pt}
  \begin{tabulary}{\textwidth}{@{}P{23mm}|P{23mm}|C|P{23mm}|P{23mm}|P{23mm}|P{20mm}@{}}
    \thead{Analysis\\(input $\rightarrow$ latents)}
    & \thead{Synthesis\\(latents $\rightarrow$ output)}
    & \thead{Hyper-Analysis\\(latents $\rightarrow$ hyperprior)}
    & \thead{Hyper-Synthesis\\(latent $\bm{\mu'}$ and $\bm{\sigma'}$)}
    & \thead{Channel-Conditional\\($\bm{\mu_i}$)}
    & \thead{Channel-Conditional\\($\bm{\sigma_i}$)}
    & \thead{Latent Residual\\Prediction}
    \\ \midrule
    Conv 5$\tines$5 c192 $\downarrow$ 2
    & Conv 5$\tines$5 c192 $\uparrow$ 2
    & Conv 3$\tines$3 c320 s1
    & Conv 5$\tines$5 c192 $\uparrow$ 2
    & Conv 3$\tines$3 c224 s1
    & Conv 3$\tines$3 c224 s1
    & Conv 3$\tines$3 c224 s1
    \\
    GDN & IGDN & ReLU & ReLU & ReLU & ReLU & ReLU
    \\
    Conv 5$\tines$5 c192 $\downarrow$ 2
    & Conv 5$\tines$5 c192 $\uparrow$ 2
    & Conv 5$\tines$5 c256 $\downarrow$ 2
    & Conv 5$\tines$5 c256 $\uparrow$ 2
    & Conv 3$\tines$3 c128 s1
    & Conv 3$\tines$3 c128 s1
    & Conv 3$\tines$3 c128 s1
    \\
    GDN & IGDN & ReLU & ReLU & ReLU & ReLU & ReLU
    \\
    Conv 5$\tines$5 c192 $\downarrow$ 2
    & Conv 5$\tines$5 c192 $\uparrow$ 2
    & Conv 5$\tines$5 c192 $\downarrow$ 2
    & Conv 3$\tines$3 c320 s1
    & Conv 3$\tines$3 c32 s1
    & Conv 3$\tines$3 c32 s1
    & Conv 3$\tines$3 c32 s1
    \\
    GDN & IGDN & & ReLU & & Exp &
    \\
    Conv 5$\tines$5 c320 $\downarrow$ 2
    & Conv 5$\tines$5 c3 $\uparrow$ 2
    & & & &
    \\
  \end{tabulary}
  \caption{Each column corresponds to a transform in the model, and each row corresponds to a layer in the transform. Convolutional layers are specified with the ``Conv'' prefix followed by the kernel size, number of channels, and up/downscaling stride where ``$\downarrow$'' represents strided convolution (downscaling), ``$\uparrow$'' represents transposed convolution (upscaling), and ``s1'' represents a stride of one. GDN stands for \underline{g}eneralized \underline{d}ivisive \underline{n}ormalization, and IGDN is inverse GDN~\cite{balle2016gdn}. The channel depths are correct for the first slice of a model with 10 slices. Since the latent tensor (the output of the analysis transform) has 320 channels, each slice has $\frac{320}{10}=32$ channels, which sets the output depth of the channel-conditional (CC) and latent residual prediction (LRP) transforms. For the remaining slices, the intermediate convolutions in the CC and LRP transforms will use larger values since the input is larger ($320 + 32 * (n-1)$ channels for slice $n$). See the text for additional details.}
  \label{table:layers}
  \vspace{4mm}
\end{table*}

Figure~\ref{fig:arch} provides a high-level overview of the network architecture for the channel-conditional model. Details about the individual transforms and layer configurations are missing due to space constraints. To facilitate reproducibility, Table~\ref{table:layers} provides detailed layer specifications for all of the transforms in a model with 10 slices.

The channel-conditional (CC) and latent residual prediction (LRP) transforms are trained separately for each slice. In each case, the output depth will be the same: \texttt{latent\_depth / num\_slices}\footnote{Typically, we use architectures where \texttt{num\_slices} evenly divides \texttt{latent\_depth}. If it doesn't, all slices have \texttt{floor(latent\_depth / num\_slices}) except for the final slice, which is set to the remaining number of channels: \texttt{slice\_depth$_{n}$ = latent\_depth - $\sum_{i=1}^{n-1}$ slice\_depth$_i$} for a model with $n$ slices.}. The input depth, however, will vary since the input to later slices include the concatenation of all previous slices. In our example, the input to slice$_1$ is 320, the input to slice$_2$ is 352, and the input to slice$_{10}$ is 608 ($320 + 32 \times 9$). To account for the different input depths, each CC and LRP transform is programmatically defined to linearly interpolate between the input and the output depth. For example, the tenth slice will have depths: 416, 224 and 32. Finally, note that the LRP transform includes the decoded values from the current block, whereas the CC transforms that predict $\mu_i$ and $\sigma_i$ values do not since it's not yet available. The input depth for the LRP transforms are thus larger than the input depths for the CC transforms by slice$_i$ channels, \ie 32 extra channels in the example model.

At low and moderate bit rates, we found that a channel depth of 320 in the latent tensor (the output of the \textit{analysis transform}) yielded good rate-distortion performance. For high bit rates, typically above 2.0 bpp on the Kodak image set~\cite{kodak}, a larger bottleneck boosts RD performance. For all reported results, we used 512 channels for such high bit rate models.

Finally, we use a simplified version of generalized divisive normalization (GDN)~\cite{balle2016gdn} where $\alpha_{ij}$ and $\varepsilon_i$ are both set to 1.0. The full formula for GDN is:

\begin{equation}
\label{eq:gdn}
z_i = \frac {x_i} {\bigl( \beta_i + \sum_j \gamma_{ij} \, |x_j|^{\alpha_{ij}} \bigr)^{\varepsilon_i}}
\end{equation}

\noindent
where $x_{i}$ and $z_{i}$ denote the input and output vectors, respectively, $ \alpha_{ij}$, $\beta_{i}$, $\gamma_{ij}$, and $\varepsilon_{i}$ represent trainable parameters, and $i$, $j$ represent channel indices. By fixing $\alpha_{ij}$ and $\varepsilon_i$ to 1.0, the simplified formula becomes:

\begin{equation}
\label{eq:1dn}
z_i = \frac {x_i} {\beta_i + \sum_j \gamma_{ij} \, |x_j|}
\end{equation}

\noindent
This change leads to slightly faster and more stable training without reducing RD performance~\cite{johnston2019}.

\section{Training Details}

For the experimental results in this paper, all models were trained for 5,000,000 steps using the Adam optimizer~\cite{KiBa15} with $\beta_1$ = 0.9, $\beta_2$ = 0.999, and $\epsilon$ = 1e-8. The learning rate started at 1e-4 and dropped to 3e-5 at 3M steps, 1e-5 at 3.6M steps, 3e-6 at 4.2M steps, and 1e-6 at 4.8M steps. Later experimentation found that a better learning rate schedule would improve RD performance by 1-2\%, even when the training duration was reduced to 4M total steps.

In addition to adjusting the learning rate, the rate-distortion trade-off parameter, $\lambda$, is also adjusted. For all experiments presented here, a model targeting $R+\lambda \cdot D$ is trained using $2 \cdot \lambda$ for the first 2.5M steps (half of the total training time). The loss function is then adjusted to use the target $\lambda$. Training with a higher $\lambda$ encourages lower distortion and thus a higher bit rate. This appears to help low bit rate models avoid a sub-optimal entropy model, but more analysis and experimentation is needed to understand exactly why this occurs and how to optimally adjust $\lambda$ during training.

All models are trained on the same images modulo random shuffling and patch extraction. The image set is made up of nearly 2M web images filtered for resolution and compression quality. The models are trained using a batch size of eight with $256 \times 256$ patches randomly cropped from the input images after random downscaling. The downscaling is useful both to reduce pre-existing compression artifacts in the training data and to discourage overfitting to a particular scale. We found that smaller patches reduced RD performance, while larger patches provided little benefit relative to slower training time.

\section{Additional Experiments}

\subsection{Partial Channel Conditioning}

\begin{figure}[tb]
  \centering
  \includegraphics[width=\columnwidth]{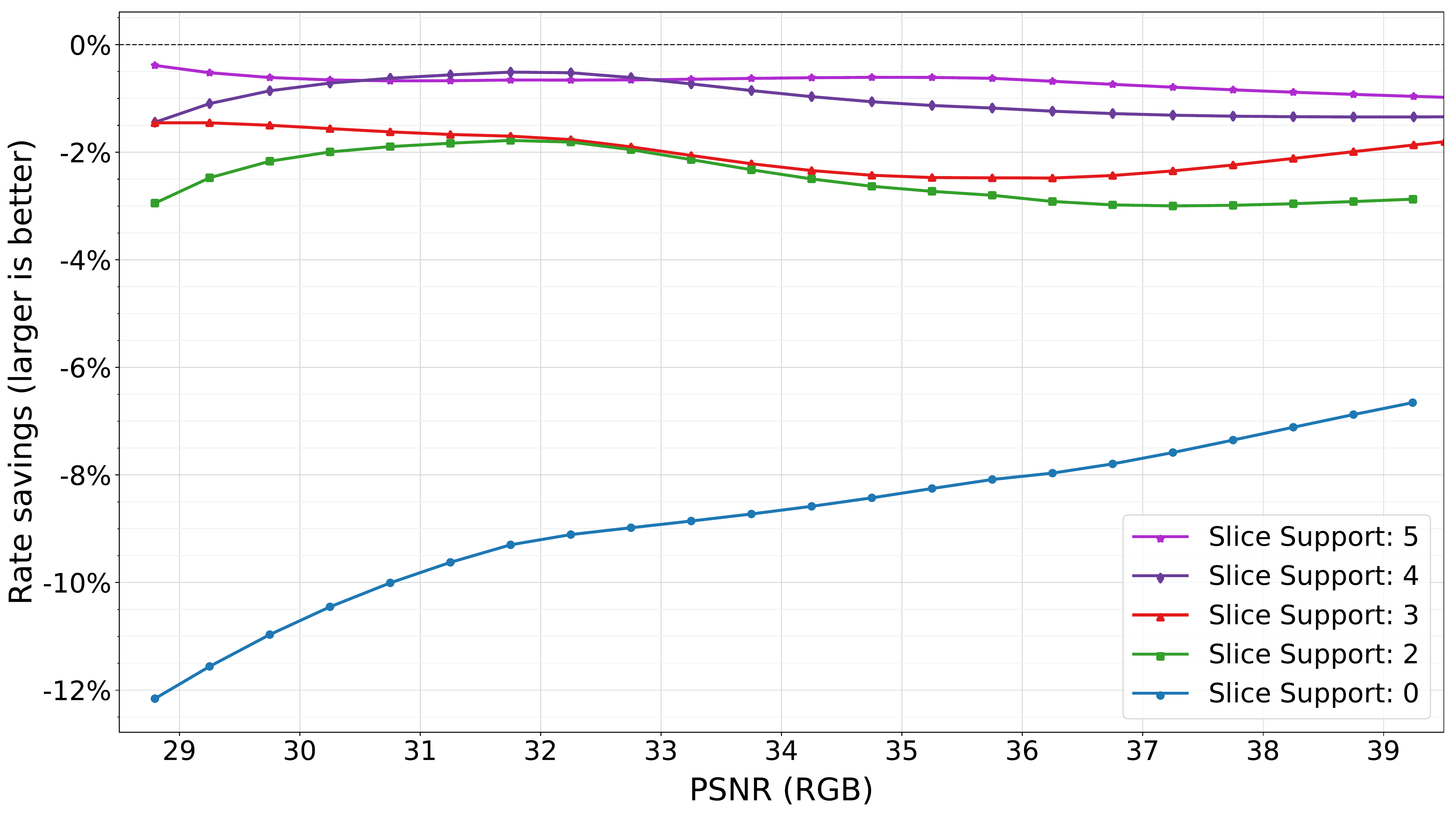}
  \shrinkcaption
  \caption{Typically, the entropy parameters for each slice are conditioned on all previous slices. This figure shows how much extra space is needed (negative rate ``savings'') as the number of supporting slices is reduced for a 10 slice model. We see that there is relatively little penalty even down to two slices, but RD performance is much worse for zero slices.}
  \label{fig:partial-slice-support}
\end{figure}

In all of the channel-conditional models explored in this paper, each slice is conditioned on all previous slices. Thus, for example, slice$_{10}$ is conditioned on nine previous slices (slice$_1$--\,slice$_9$), which leads to a relatively large input depth for the later slices and thus slower models. Figure~\ref{fig:partial-slice-support} shows how RD performance degrades as slices are conditioned on fewer previous slices. For example, in a 10 slice model, if the slice support is five, the final slice will only be conditioned on the first five slices and will be conditionally independent from slice$_6$--\,slice$_9$. Reducing the slice support decreases the size of the channel-conditional transform for later slices and creates more opportunity for parallel calculations, both of which improve runtime.

In addition to exploring the effect of conditioning on a limited number of early slices (slice$_1$--\,slice$_N$), we also explored conditioning on the \textit{previous} N slices. For example, if conditioning on five slices, slice$_7$ would be conditioned on slice$_2$--slice$_6$. This approach led to slightly worse RD performance and provides fewer opportunities for parallelization compared to always conditioning on the first N slices.

Figure~\ref{fig:partial-slice-support} shows that reducing the slice support does reduce RD performance, but the reduction is relatively small. Further research is needed to fully understand how the latent representation in the early slices changes, but our theory is that most of the benefit of channel-conditioning comes from a relatively small amount of high-level information. Models that condition on fewer slices learn to represent this information in the early slices, thus preserving the overall effectiveness of the model.

\subsection{Smaller Hyperprior and $\mu', \sigma'$ Tensors}

\begin{figure}[tb]
  \centering
  \includegraphics[width=\columnwidth]{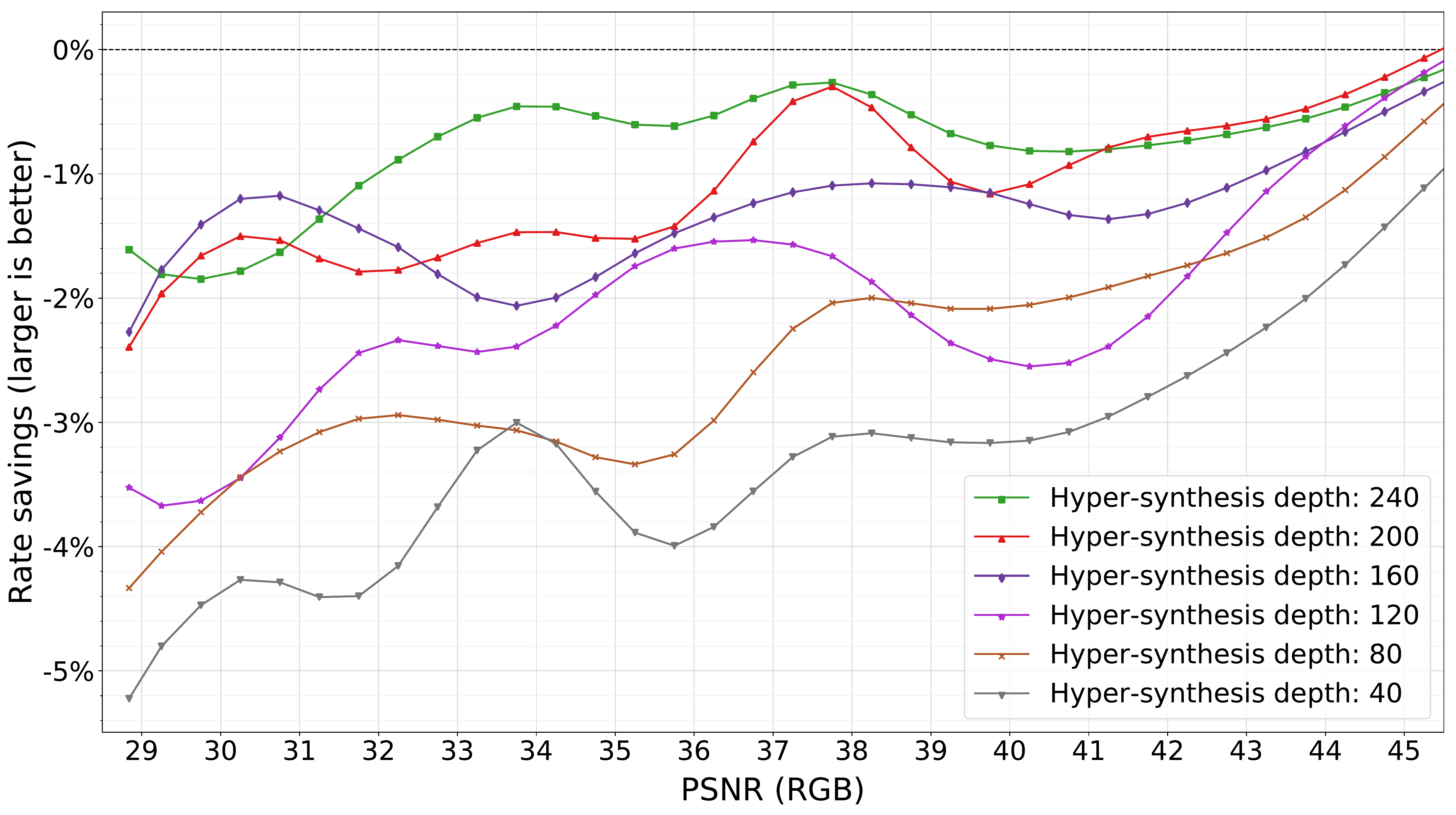}
  \shrinkcaption
  \caption{Our standard model architecture uses 320 channels in the output of the hyper-synthesis transforms. This figure shows how compressed images get larger (negative rate ``savings'') when the number of hyper-synthesis output channels decreases. We see that reducing the depth from 320 to 240 has a very small effect (file size grows by less than 2\%), and even very small latent tensors (40 channels) are only \textasciitilde 5\% larger in the worst case.}
  \label{fig:smaller-hyper-depth}
\end{figure}

In channel-conditional models, all slices are conditioned on previous slices as well as a latent tensor predicted from the hyperprior ($\mu'$ and $\sigma'$ in Figure~\ref{fig:arch}). Each slice has its own transform for predicting the mean and scale values in that slice ($\mu_i$ and $\sigma_i$), and the size and speed of the transform depends on the size of $\mu'$ and $\sigma'$. All of the models in the main paper use 320 channels for these tensors, which is relatively large. We therefore explored the effect on RD performance as the depth of these tensors is reduced (see Figure~\ref{fig:smaller-hyper-depth}). The experiment confirms that larger tensors help, at least up to 320 channels, but the RD penalty for shrinking the tensor is minimal (\textasciitilde 2\% in the worst case) down to a depth of 160 channels.

Further experiments explored the effect of shrinking the hyper-analysis transform. Our typical model transforms a 320 channel latent into a 192 channel hyperprior (see the \textit{Hyper-Analysis} column in Table~\ref{table:layers}). Empirically, we found that shrinking the hyper-analysis transform from layers with depths of [320, 256, 192] to depths of [128, 86, 64] had only a small impact on RD performance.

\section{Rate-Distortion Comparisons}

\begin{figure*}[tb]
  \centering
  \includegraphics[width=0.95\linewidth]{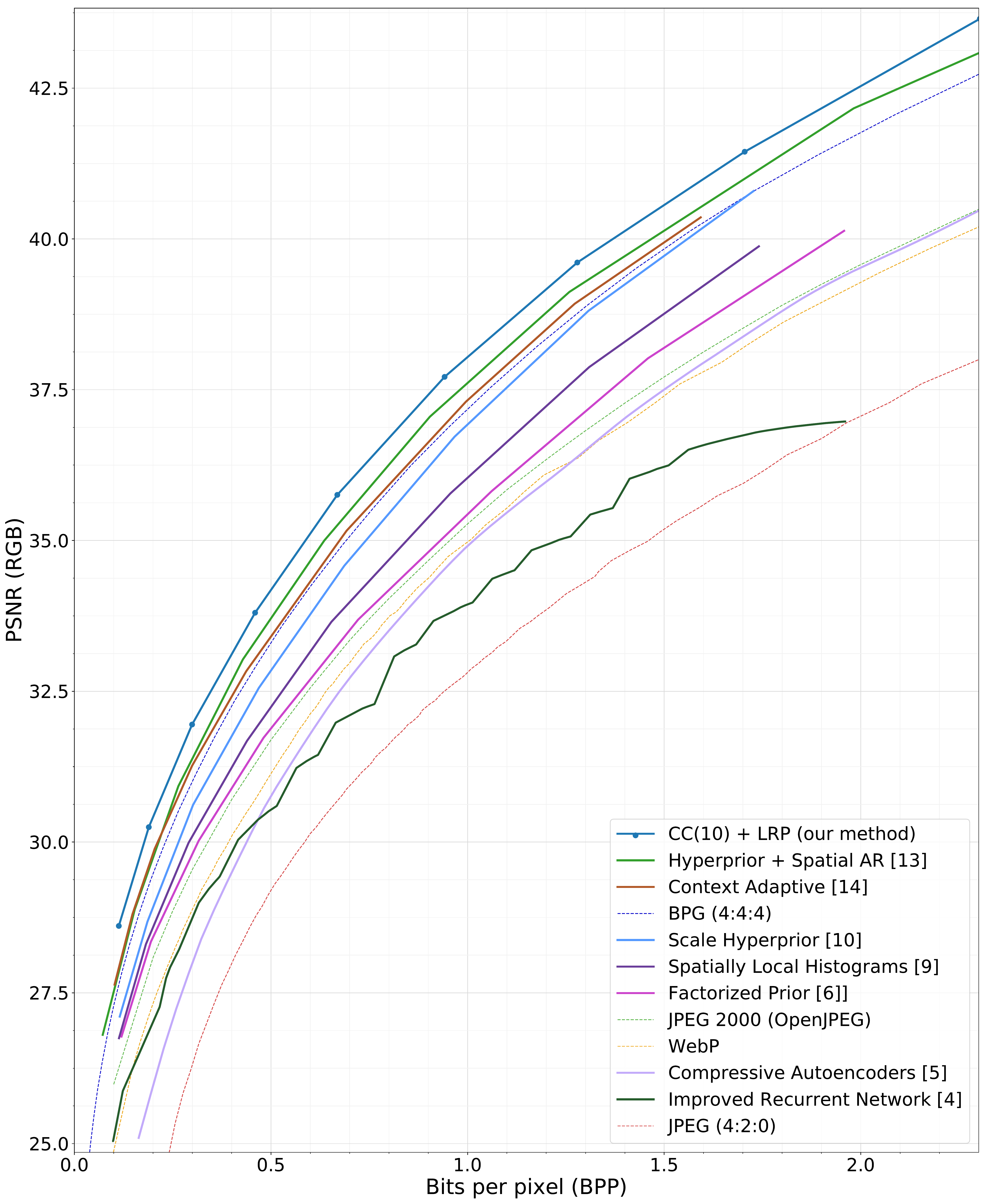}
  \shrinkcaption
  \caption{This graph of rate-distortion (RD) curves shows how our model outperforms a wide range of existing learning-based and standard codecs on the Kodak image set~\cite{kodak} using PSNR as the image quality metric.}
  \label{fig:big-rd}
\end{figure*}

Figure~\ref{fig:kodak-rd} in the main paper provides a rate-distortion comparison between our method and a small set of recent learning-based methods as well as several standard methods. Figure~\ref{fig:big-rd} extends this comparison by including many more compression methods and providing a larger graph to aid readability.

\begin{figure}[tb]
  \centering
  \includegraphics[width=\linewidth]{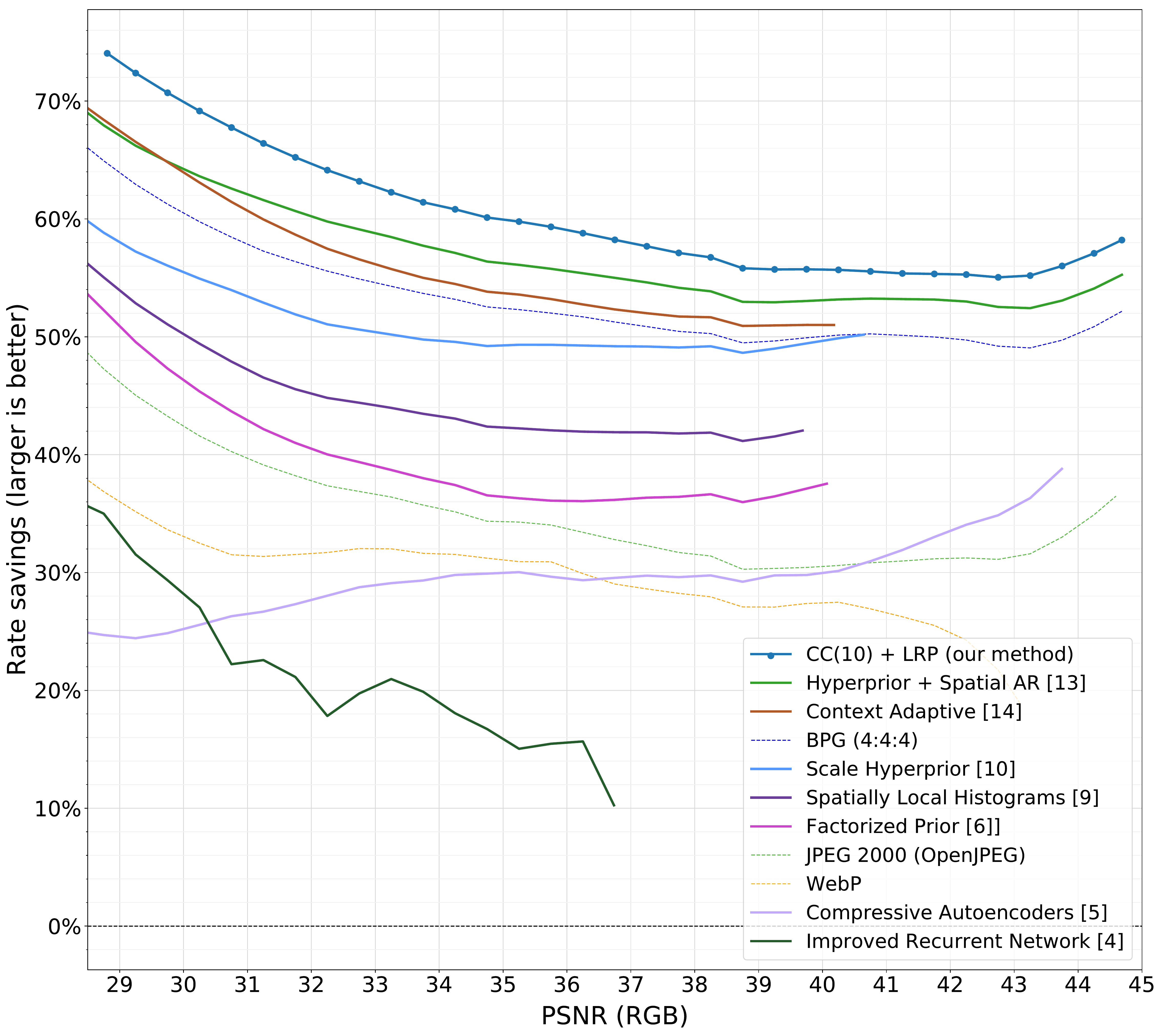}
  \shrinkcaption
  \caption{This graph shows rate savings (higher is better) compared to JPEG (4:2:0) compression. It shows that our model outperforms a wide range of existing learning-based and standard codecs on the Kodak image set~\cite{kodak} using PSNR as the image quality metric.}
  \label{fig:big-bdbins}
\end{figure}

Figure~\ref{fig:big-bdbins} shows the same data in a format that highlights the rate savings of different compression methods relative to JPEG (4:2:0). In this case, the curves represent the file size reduction as a percentage of the size of a JPEG encoding with equivalent PSNR. Larger values imply higher savings, and the graph shows that our method can shrink images by more than 70\% compared to JPEG at lower quality levels and by more than 55\% at all quality levels.

\section{Reconstructed Images}

\begin{figure*}[t]
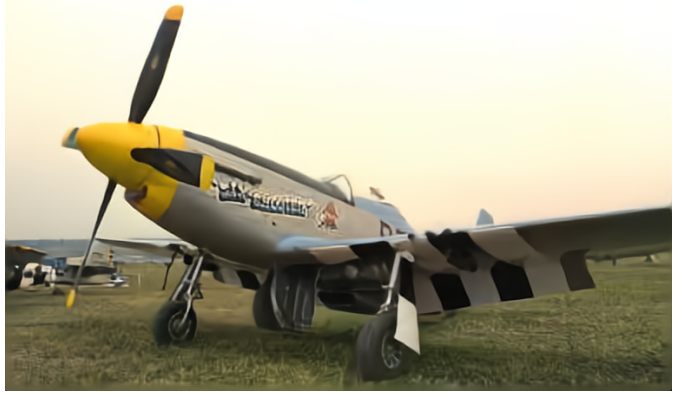
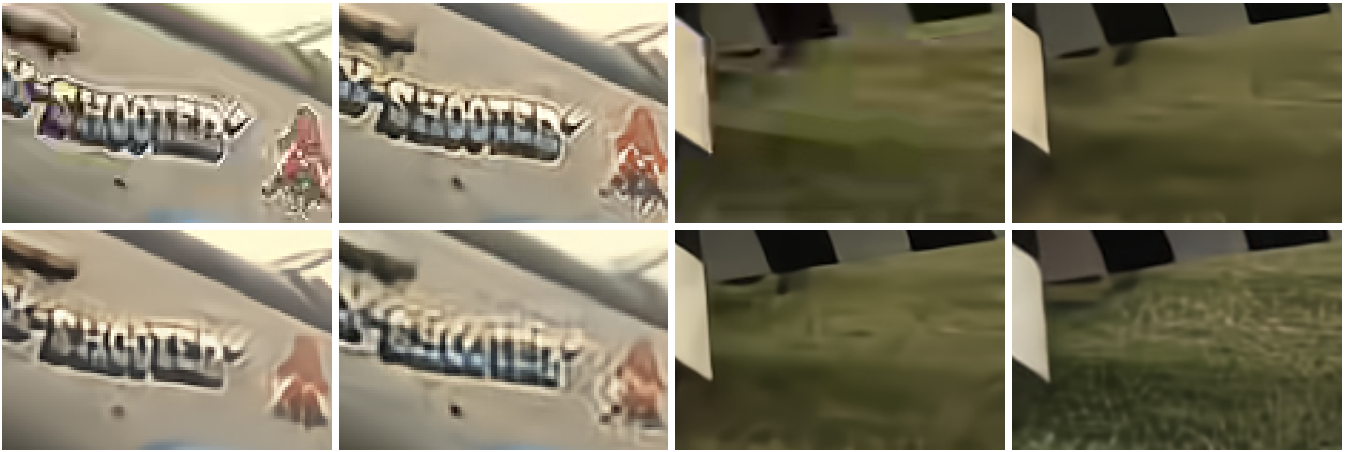

\centering

\newlength{\margin}
\newlength{\bigwidth}
\newlength{\smallwidth}

\setlength{\margin}{0.006\linewidth}
\setlength{\bigwidth}{(\linewidth - \margin) / 2}
\setlength{\smallwidth}{(\linewidth - 3\margin) / 4}

\newcommand{\trimAleft}{0.35}
\newcommand{\trimAright}{0.5}
\newcommand{\trimAtop}{0.5}
\newcommand{\trimAbottom}{0.35}
\newcommand{\boxAright}{1 - \trimAright}
\newcommand{\boxAtop}{1 - \trimAtop}

\newcommand{\trimBleft}{0.6}
\newcommand{\trimBright}{0.2}
\newcommand{\trimBtop}{0.78}
\newcommand{\trimBbottom}{0.02}
\newcommand{\boxBright}{1 - \trimBright}
\newcommand{\boxBtop}{1 - \trimBtop}

\newcommand{\imageA}{{figures/reconstructions/kodim20-bpg-444-q42-0.06720}.png}
\newcommand{\imageB}{{figures/reconstructions/kodim20-cc8-mse-0.0662}.png}
\newcommand{\imageC}{{figures/reconstructions/kodim20-cc8-l1-0.0698}.png}
\newcommand{\imageD}{{figures/reconstructions/kodim20-cc8-msssim-0.0626}.png}

\begin{tikzpicture}
    \node[anchor=south west,inner sep=0, outer sep=0] (bigA) at (0,0) {\includegraphics[width=\bigwidth]{\imageA}};
    \begin{scope}[x={(bigA.south east)},y={(bigA.north west)}]
        \draw[red,thick,rounded corners](\trimAleft,\trimAbottom) rectangle (\boxAright,\boxAtop);
        \draw[green,thick,rounded corners](\trimBleft,\trimBbottom) rectangle (\boxBright,\boxBtop);
    \end{scope}
    \node[anchor=north west] at (bigA.north west) {\textsf{BPG @ 0.0672 bpp}};
    
    \node[anchor=south west,inner sep=0, outer sep=0, right=\margin of bigA] (bigB) {\includegraphics[width=\bigwidth]{\imageB}};
    \node[anchor=north west] at (bigB.north west) {\textsf{CC(8) opt. for MSE @ 0.0662 bpp}};
    
    \node[anchor=south west,inner sep=0, outer sep=0, below=\margin of bigA] (bigC) {\includegraphics[width=\bigwidth]{\imageC}};
    \node[anchor=north west] at (bigC.north west) {\textsf{CC(8) opt. for L1 @ 0.0698 bpp}};
    
    \node[anchor=south west,inner sep=0, outer sep=0, right=\margin of bigC] (bigD) {\includegraphics[width=\bigwidth]{\imageD}};
    \node[anchor=north west] at (bigD.north west) {\textsf{CC(8) opt. for MS-SSIM @ 0.0626 bpp}};
    
    \node[inner sep=0, outer sep=0, below=\margin of bigC.south west, anchor=north west] (zoomA1)
    {\adjincludegraphics[width=\smallwidth,trim={\trimAleft\width} {\trimAbottom\height} {\trimAright\width} {\trimAtop\height}, clip]{\imageA}};
    
    \node[anchor=south west,inner sep=0, outer sep=0, right=\margin of zoomA1] (zoomB1)
    {\adjincludegraphics[width=\smallwidth,trim={\trimAleft\width} {\trimAbottom\height} {\trimAright\width} {\trimAtop\height}, clip]{\imageB}};
    
    \node[anchor=south west,inner sep=0, outer sep=0, below=\margin of zoomA1] (zoomC1)
    {\adjincludegraphics[width=\smallwidth,trim={\trimAleft\width} {\trimAbottom\height} {\trimAright\width} {\trimAtop\height}, clip]{\imageC}};
    
    \node[anchor=south west,inner sep=0, outer sep=0, right=\margin of zoomC1] (zoomD1)
    {\adjincludegraphics[width=\smallwidth,trim={\trimAleft\width} {\trimAbottom\height} {\trimAright\width} {\trimAtop\height}, clip]{\imageD}};
    
    \node[inner sep=0, outer sep=0, below=\margin of bigD.south west, anchor=north west] (zoomA2)
    {\adjincludegraphics[width=\smallwidth,trim={\trimBleft\width} {\trimBbottom\height} {\trimBright\width} {\trimBtop\height}, clip]{\imageA}};
    
    \node[anchor=south west,inner sep=0, outer sep=0, right=\margin of zoomA2] (zoomB2)
    {\adjincludegraphics[width=\smallwidth,trim={\trimBleft\width} {\trimBbottom\height} {\trimBright\width} {\trimBtop\height}, clip]{\imageB}};
    
    \node[anchor=south west,inner sep=0, outer sep=0, below=\margin of zoomA2] (zoomC2)
    {\adjincludegraphics[width=\smallwidth,trim={\trimBleft\width} {\trimBbottom\height} {\trimBright\width} {\trimBtop\height}, clip]{\imageC}};
    
    \node[anchor=south west,inner sep=0, outer sep=0, right=\margin of zoomC2] (zoomD2)
    {\adjincludegraphics[width=\smallwidth,trim={\trimBleft\width} {\trimBbottom\height} {\trimBright\width} {\trimBtop\height}, clip]{\imageD}};
    
\end{tikzpicture}

\caption{The top four images are reconstructions of kodim20 from the Kodak image set~\cite{kodak} from four different codecs after significant compression (roughly 360x). Below the full-size images are two crops. The first (red box) highlights how optimizing for MSE best maintains legible text, while the second (green box) shows that only the model optimized for MS-SSIM maintains any reasonable texture in the grass.}
\label{fig:kodim20}
\end{figure*}

\begin{figure*}[t]
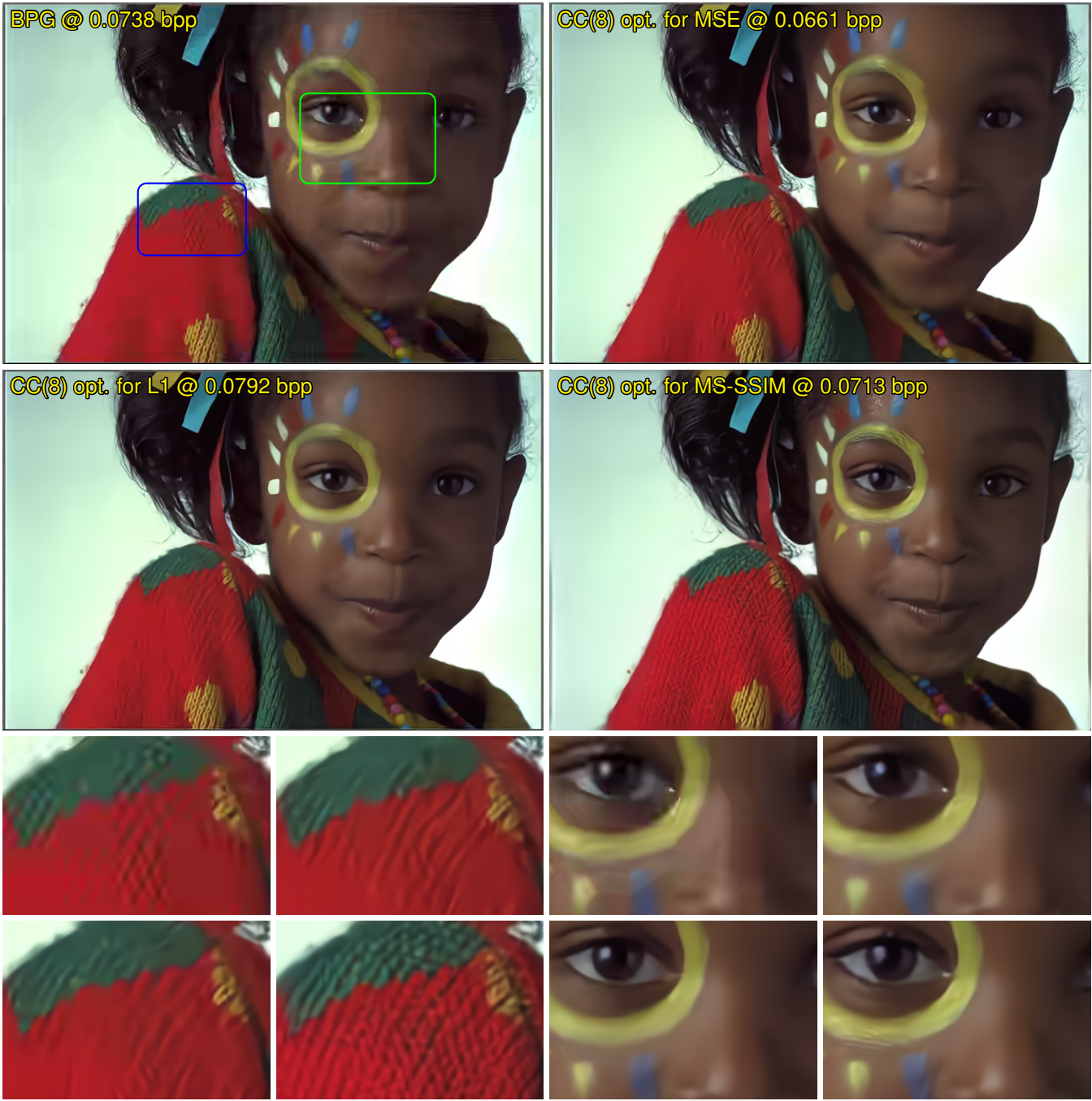

\centering

\setlength{\margin}{0.006\linewidth}
\setlength{\bigwidth}{(\linewidth - \margin) / 2}
\setlength{\smallwidth}{(\linewidth - 3\margin) / 4}

\newcommand{\trimAleft}{0.25}
\newcommand{\trimAright}{0.55}
\newcommand{\trimAtop}{0.5}
\newcommand{\trimAbottom}{0.3}
\newcommand{\boxAright}{1 - \trimAright}
\newcommand{\boxAtop}{1 - \trimAtop}

\newcommand{\trimBleft}{0.55}
\newcommand{\trimBright}{0.2}
\newcommand{\trimBtop}{0.25}
\newcommand{\trimBbottom}{0.5}
\newcommand{\boxBright}{1 - \trimBright}
\newcommand{\boxBtop}{1 - \trimBtop}

\newcommand{\imageA}{{figures/reconstructions/kodim15-bpg-444-q41-0.07381}.png}
\newcommand{\imageB}{{figures/reconstructions/kodim15-cc8-mse-0.0661}.png}
\newcommand{\imageC}{{figures/reconstructions/kodim15-cc8-l1-0.0792}.png}
\newcommand{\imageD}{{figures/reconstructions/kodim15-cc8-msssim-0.0713}.png}

\begin{tikzpicture}
    \node[anchor=south west,inner sep=0, outer sep=0] (bigA) at (0,0) {\includegraphics[width=\bigwidth]{\imageA}};
    \begin{scope}[x={(bigA.south east)},y={(bigA.north west)}]
        \draw[blue,thick,rounded corners](\trimAleft,\trimAbottom) rectangle (\boxAright,\boxAtop);
        \draw[green,thick,rounded corners](\trimBleft,\trimBbottom) rectangle (\boxBright,\boxBtop);
    \end{scope}
    \node[anchor=north west,text=yellow] at (bigA.north west) {\contour{black}{\textcolor{yellow}{\textsf{BPG @ 0.0738 bpp}}}};
    
    \node[anchor=south west,inner sep=0, outer sep=0, right=\margin of bigA] (bigB) {\includegraphics[width=\bigwidth]{\imageB}};
    \node[anchor=north west] at (bigB.north west) {\contour{black}{\textcolor{yellow}{\textsf{CC(8) opt. for MSE @ 0.0661 bpp}}}};
    
    \node[anchor=south west,inner sep=0, outer sep=0, below=\margin of bigA] (bigC) {\includegraphics[width=\bigwidth]{\imageC}};
    \node[anchor=north west,text=yellow] at (bigC.north west) {\contour{black}{\textcolor{yellow}{\textsf{CC(8) opt. for L1 @ 0.0792 bpp}}}};
    
    \node[anchor=south west,inner sep=0, outer sep=0, right=\margin of bigC] (bigD) {\includegraphics[width=\bigwidth]{\imageD}};
    \node[anchor=north west,text=yellow] at (bigD.north west) {\contour{black}{\textcolor{yellow}{\textsf{CC(8) opt. for MS-SSIM @ 0.0713 bpp}}}};
    
    \node[inner sep=0, outer sep=0, below=\margin of bigC.south west, anchor=north west] (zoomA1)
    {\adjincludegraphics[width=\smallwidth,trim={\trimAleft\width} {\trimAbottom\height} {\trimAright\width} {\trimAtop\height}, clip]{\imageA}};
    
    \node[anchor=south west,inner sep=0, outer sep=0, right=\margin of zoomA1] (zoomB1)
    {\adjincludegraphics[width=\smallwidth,trim={\trimAleft\width} {\trimAbottom\height} {\trimAright\width} {\trimAtop\height}, clip]{\imageB}};
    
    \node[anchor=south west,inner sep=0, outer sep=0, below=\margin of zoomA1] (zoomC1)
    {\adjincludegraphics[width=\smallwidth,trim={\trimAleft\width} {\trimAbottom\height} {\trimAright\width} {\trimAtop\height}, clip]{\imageC}};
    
    \node[anchor=south west,inner sep=0, outer sep=0, right=\margin of zoomC1] (zoomD1)
    {\adjincludegraphics[width=\smallwidth,trim={\trimAleft\width} {\trimAbottom\height} {\trimAright\width} {\trimAtop\height}, clip]{\imageD}};
    
    \node[inner sep=0, outer sep=0, below=\margin of bigD.south west, anchor=north west] (zoomA2)
    {\adjincludegraphics[width=\smallwidth,trim={\trimBleft\width} {\trimBbottom\height} {\trimBright\width} {\trimBtop\height}, clip]{\imageA}};
    
    \node[anchor=south west,inner sep=0, outer sep=0, right=\margin of zoomA2] (zoomB2)
    {\adjincludegraphics[width=\smallwidth,trim={\trimBleft\width} {\trimBbottom\height} {\trimBright\width} {\trimBtop\height}, clip]{\imageB}};
    
    \node[anchor=south west,inner sep=0, outer sep=0, below=\margin of zoomA2] (zoomC2)
    {\adjincludegraphics[width=\smallwidth,trim={\trimBleft\width} {\trimBbottom\height} {\trimBright\width} {\trimBtop\height}, clip]{\imageC}};
    
    \node[anchor=south west,inner sep=0, outer sep=0, right=\margin of zoomC2] (zoomD2)
    {\adjincludegraphics[width=\smallwidth,trim={\trimBleft\width} {\trimBbottom\height} {\trimBright\width} {\trimBtop\height}, clip]{\imageD}};
    
\end{tikzpicture}

\caption{The top four images are reconstructions of kodim15 from the Kodak image set~\cite{kodak} from four different codecs after significant compression (roughly 330x). Below the full-size images are two crops. The first (blue box) highlights the additional texture that is maintained by the channel-conditional model optimized for MS-SSIM. The second crop (green box) highlights the geometric distortions introduced by BPG. All four methods produce overly smooth skin at these very low bit rates.}
\label{fig:kodim15}
\end{figure*}

\begin{figure*}[t]
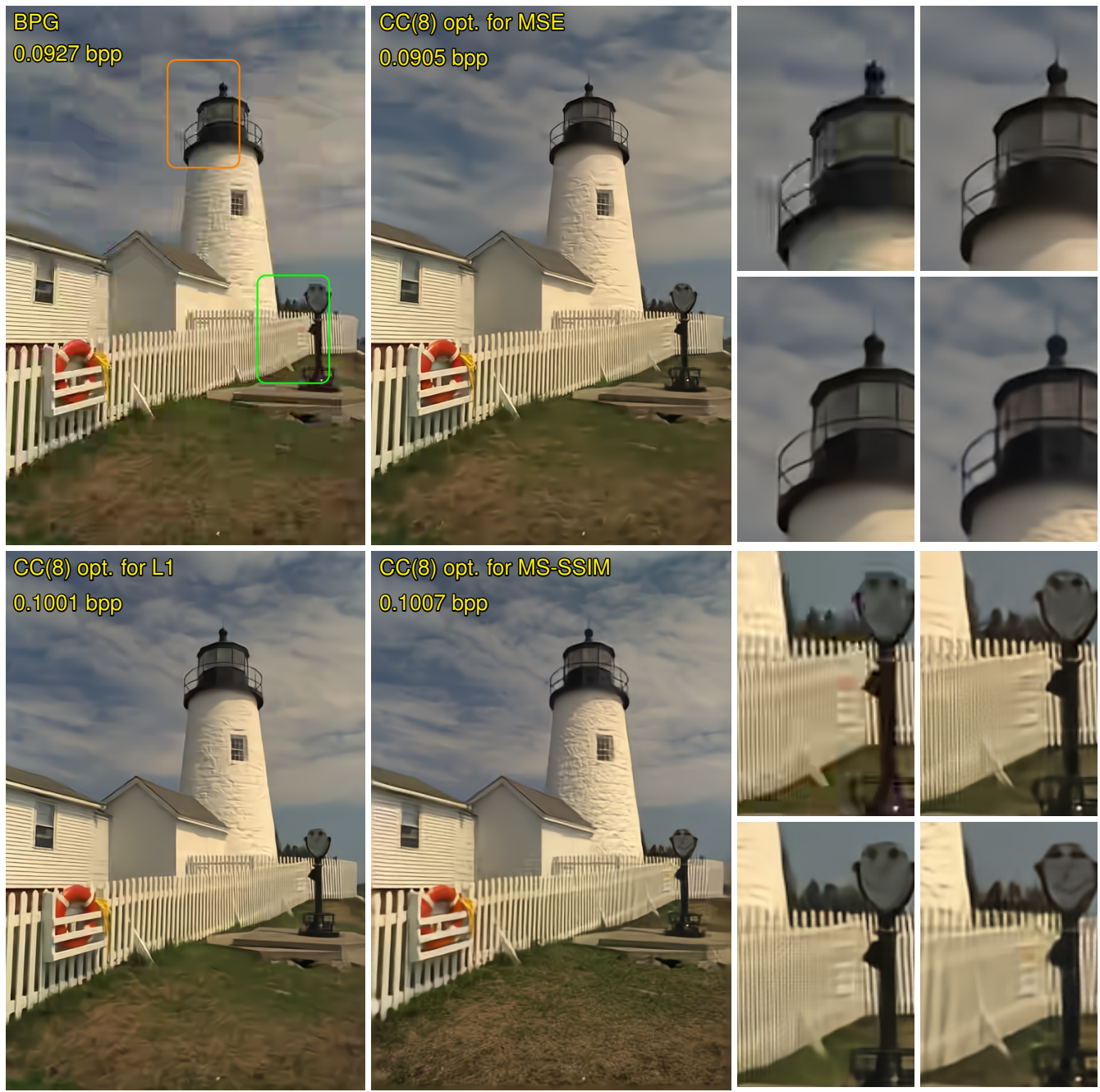

\centering

\setlength{\margin}{0.006\linewidth}
\setlength{\bigwidth}{(0.66\linewidth - \margin) / 2}
\setlength{\smallwidth}{(0.66\linewidth - 3\margin) / 4}

\newcommand{\trimAleft}{0.45}
\newcommand{\trimAright}{0.35}
\newcommand{\trimAtop}{0.1}
\newcommand{\trimAbottom}{0.7}
\newcommand{\boxAright}{1 - \trimAright}
\newcommand{\boxAtop}{1 - \trimAtop}

\newcommand{\trimBleft}{0.7}
\newcommand{\trimBright}{0.1}
\newcommand{\trimBtop}{0.5}
\newcommand{\trimBbottom}{0.3}
\newcommand{\boxBright}{1 - \trimBright}
\newcommand{\boxBtop}{1 - \trimBtop}

\newcommand{\imageA}{{figures/reconstructions/kodim19-bpg-444-q42-0.09273}.png}
\newcommand{\imageB}{{figures/reconstructions/kodim19-cc8-mse-0.0905}.png}
\newcommand{\imageC}{{figures/reconstructions/kodim19-cc8-l1-0.1001}.png}
\newcommand{\imageD}{{figures/reconstructions/kodim19-cc8-msssim-0.1007}.png}

\begin{tikzpicture}
    \node[anchor=south west,inner sep=0, outer sep=0] (bigA) at (0,0) {\includegraphics[width=\bigwidth]{\imageA}};
    \begin{scope}[x={(bigA.south east)},y={(bigA.north west)}]
        \draw[orange,thick,rounded corners](\trimAleft,\trimAbottom) rectangle (\boxAright,\boxAtop);
        \draw[green,thick,rounded corners](\trimBleft,\trimBbottom) rectangle (\boxBright,\boxBtop);
    \end{scope}
    \node[anchor=north west] (line1) at (bigA.north west) {\contour{black}{\textcolor{yellow}{\textsf{BPG}}}};
    \node[anchor=north west] at (line1.south west) {\contour{black}{\textcolor{yellow}{\textsf{0.0927 bpp}}}};
    
    \node[anchor=south west,inner sep=0, outer sep=0, right=\margin of bigA] (bigB) {\includegraphics[width=\bigwidth]{\imageB}};
    \node[anchor=north west] (line1) at (bigB.north west) {\contour{black}{\textcolor{yellow}{\textsf{CC(8) opt. for MSE}}}};
    \node[anchor=north west] at (line1.south west) {\contour{black}{\textcolor{yellow}{\textsf{0.0905 bpp}}}};

    \node[anchor=south west,inner sep=0, outer sep=0, below=\margin of bigA] (bigC) {\includegraphics[width=\bigwidth]{\imageC}};
    \node[anchor=north west] (line1) at (bigC.north west) {\contour{black}{\textcolor{yellow}{\textsf{CC(8) opt. for L1}}}};
    \node[anchor=north west] at (line1.south west) {\contour{black}{\textcolor{yellow}{\textsf{0.1001 bpp}}}};
    
    \node[anchor=south west,inner sep=0, outer sep=0, right=\margin of bigC] (bigD) {\includegraphics[width=\bigwidth]{\imageD}};
    \node[anchor=north west] (line1) at (bigD.north west) {\contour{black}{\textcolor{yellow}{\textsf{CC(8) opt. for MS-SSIM}}}};
    \node[anchor=north west] at (line1.south west) {\contour{black}{\textcolor{yellow}{\textsf{0.1007 bpp}}}};
    
    \node[inner sep=0, outer sep=0, right=\margin of bigB.north east, anchor=north west] (zoomA1)
    {\adjincludegraphics[width=\smallwidth,trim={\trimAleft\width} {\trimAbottom\height} {\trimAright\width} {\trimAtop\height}, clip]{\imageA}};
    
    \node[anchor=south west,inner sep=0, outer sep=0, right=\margin of zoomA1] (zoomB1)
    {\adjincludegraphics[width=\smallwidth,trim={\trimAleft\width} {\trimAbottom\height} {\trimAright\width} {\trimAtop\height}, clip]{\imageB}};
    
    \node[anchor=south west,inner sep=0, outer sep=0, below=\margin of zoomA1] (zoomC1)
    {\adjincludegraphics[width=\smallwidth,trim={\trimAleft\width} {\trimAbottom\height} {\trimAright\width} {\trimAtop\height}, clip]{\imageC}};
    
    \node[anchor=south west,inner sep=0, outer sep=0, right=\margin of zoomC1] (zoomD1)
    {\adjincludegraphics[width=\smallwidth,trim={\trimAleft\width} {\trimAbottom\height} {\trimAright\width} {\trimAtop\height}, clip]{\imageD}};
    
    \node[inner sep=0, outer sep=0, right=\margin of bigD.north east, anchor=north west] (zoomA2)
    {\adjincludegraphics[width=\smallwidth,trim={\trimBleft\width} {\trimBbottom\height} {\trimBright\width} {\trimBtop\height}, clip]{\imageA}};
    
    \node[anchor=south west,inner sep=0, outer sep=0, right=\margin of zoomA2] (zoomB2)
    {\adjincludegraphics[width=\smallwidth,trim={\trimBleft\width} {\trimBbottom\height} {\trimBright\width} {\trimBtop\height}, clip]{\imageB}};
    
    \node[anchor=south west,inner sep=0, outer sep=0, below=\margin of zoomA2] (zoomC2)
    {\adjincludegraphics[width=\smallwidth,trim={\trimBleft\width} {\trimBbottom\height} {\trimBright\width} {\trimBtop\height}, clip]{\imageC}};
    
    \node[anchor=south west,inner sep=0, outer sep=0, right=\margin of zoomC2] (zoomD2)
    {\adjincludegraphics[width=\smallwidth,trim={\trimBleft\width} {\trimBbottom\height} {\trimBright\width} {\trimBtop\height}, clip]{\imageD}};
    
\end{tikzpicture}

\caption{The left four images are reconstructions of kodim19 from the Kodak image set~\cite{kodak} from different codecs after significant compression (roughly 250x). The first crop (orange box) highlights ringing artifacts from BPG around the railing of the lighthouse, while the second crop (green box) shows how optimizing for L1 or MS-SSIM leads to artifacts and blurring in the fence. As is typical, only MS-SSIM preserves any reasonable texture in the grass, and MSE overly smooths the sky. BPG is the only codec that preserves the red color in the sign near the end of the fence, though no codec is able to preserve legibility (the sign says "Danger"). On the other hand, BPG introduces considerable blocking and geometric artifacts in the sky.}
\label{fig:kodim19}
\end{figure*}

Reconstructed images are shown in Figures~\ref{fig:kodim20},~\ref{fig:kodim15}, and~\ref{fig:kodim19} comparing BPG with our channel-conditional model optimized for three different metrics (MSE, L1, and MS-SSIM). These example images were compressed at very high rates (ranging from roughly 250x to nearly 360x compression) since high compression rates help highlight the kinds of distortions typical for each method and quality metric. For example, optimizing for MS-SSIM typically preserves texture better than MSE or L1, \eg in the grass in Figures~\ref{fig:kodim20} and~\ref{fig:kodim19} and the red sweater in Figure~\ref{fig:kodim15}. MS-SSIM performs the worst, however, on high-contrast and high-frequency content like text, as shown in Figure~\ref{fig:kodim20}, and in the receding fence in Figure~\ref{fig:kodim19}.



This paper focuses on a more effective and more efficient entropy model based on  channel-conditioning (CC) and latent residual prediction (LRP) compared to spatially autoregressive models or solely hierarchical priors. The benefits of the CC and LRP model are independent of the loss function used to quantify visual distortions. Interesting future research could look at combining our entropy model with more sophisticated image quality metrics such as perceptual metrics~\cite{chinen2018icip,zhang2018cvpr,ding2020arxiv} and adversarial loss~\cite{agustsson2018generative,mentzer2020arxiv,wu2020arxiv}.

\section{Sampling from the Compression Model}

\subsection{Samples with a Random Hyperprior}

\begin{figure*}[tb]
  \centering
  \includegraphics[width=0.33\linewidth]{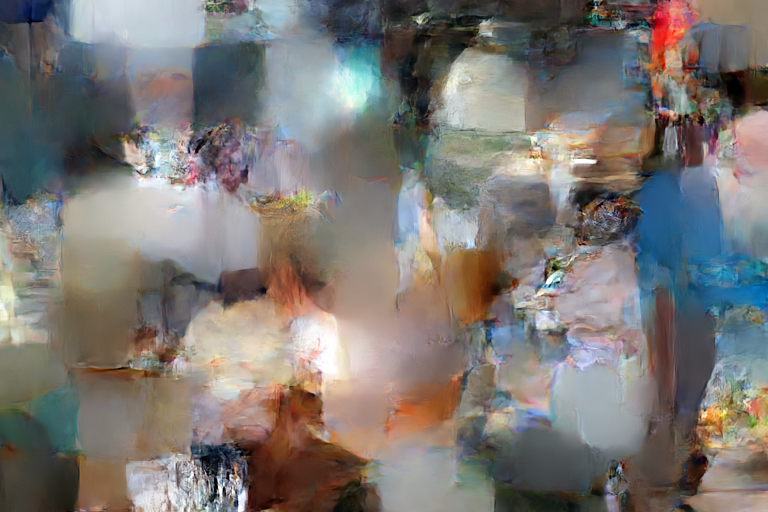}
  \hfill
  \includegraphics[width=0.33\linewidth]{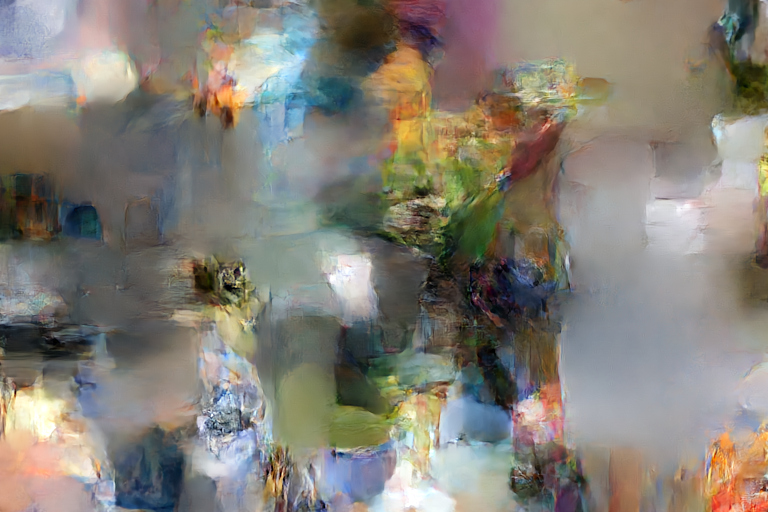}
  \hfill
  \includegraphics[width=0.33\linewidth]{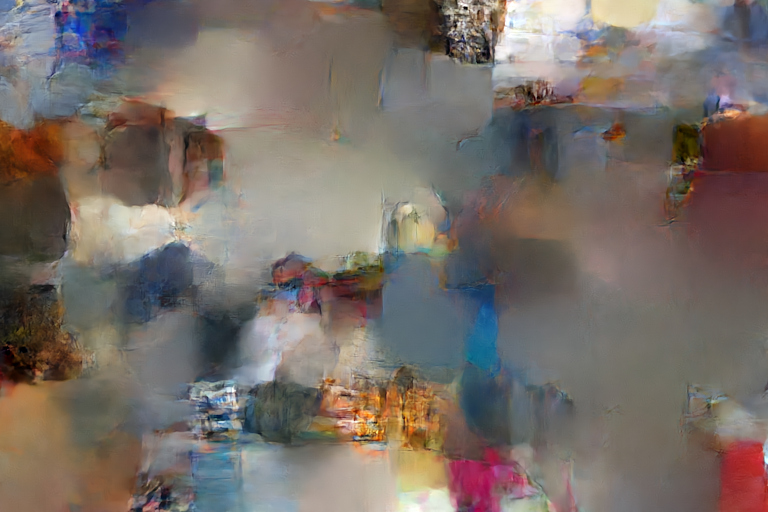}
  \shrinkcaption
  \caption{Random samples from a channel-conditional model optimized on a semantically unconstrained set of nearly two million high-resolution web images. Although the samples do show some local coherence, no obvious semantic information is visible. Contrast with typical results from GAN-based models optimized on semantically constrained image sets.}
  \label{fig:samples}
\end{figure*}

Since the compression model learns a factorized distribution over the hyperprior as well as a conditional distribution over the latents, we can treat the network as a generative model and sample random images. Doing so provides a visualization of what kind of images are typical according to the image distribution learned by the model. See Figure~\ref{fig:samples} for several examples for a CC(10) model.

The sampling process is straightforward. First, draw a sample from the factorized entropy model to generate a random hyperprior. Next, run the hyper-synthesis transform to get $\mu'$ and $\sigma'$. Then iterate over each slice of the latent tensor. For each slice$_i$, run the $\mu_i$ and $\sigma_i$ transforms to get an entropy model for the slice and draw a random sample. After all slices have been sampled, concatenate the result and run the synthesis transform to generate an RGB image.

\begin{figure*}[tb]
  \captionsetup{justification=centering}

  
  \setlength\figwidth{0.162\linewidth}
  \setlength\imagewidth{0.161\linewidth}
  \setlength\subcap{-20pt}
  
  \centering
  \begin{subfigure}[t]{\figwidth}
    \centering
    \includegraphics[width=\imagewidth]{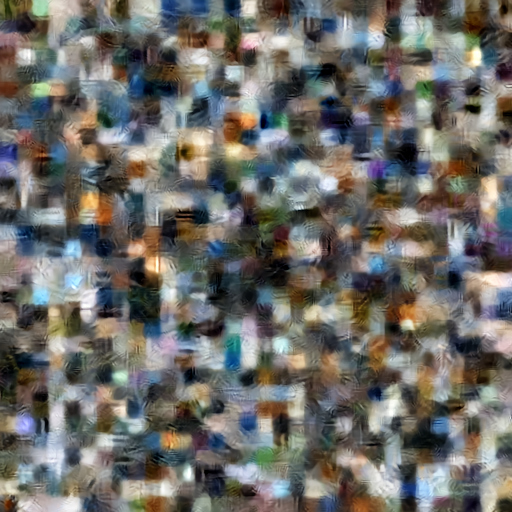}
    \vspace{\subcap}
    \caption*{Factorized Prior~\cite{balle2017iclr}}
  \end{subfigure}
  \hfill
  \begin{subfigure}[t]{\figwidth}
    \centering
    \includegraphics[width=\imagewidth]{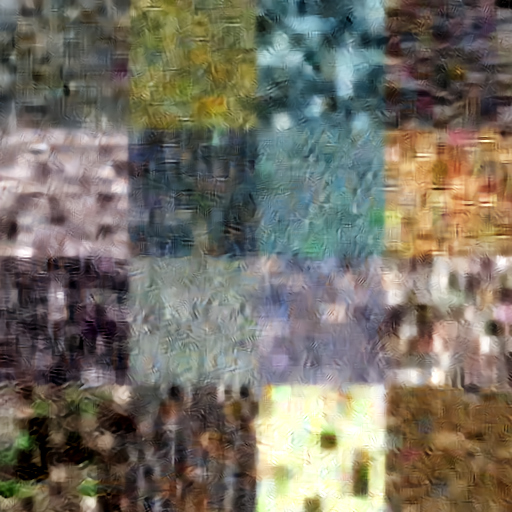}
    \vspace{\subcap}
    \caption*{Spatially Local Histograms~\cite{minnen2018icip}}
  \end{subfigure}
  \hfill
  \begin{subfigure}[t]{\figwidth}
    \centering
    \includegraphics[width=\imagewidth]{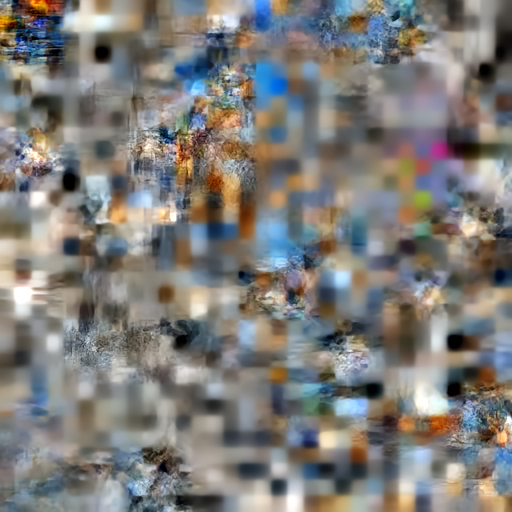}
    \vspace{\subcap}
    \caption*{Hyperprior\\(Scale-only)~\cite{balle2018iclr}}
  \end{subfigure}
  \hfill
  \begin{subfigure}[t]{\figwidth}
    \centering
    \includegraphics[width=\imagewidth]{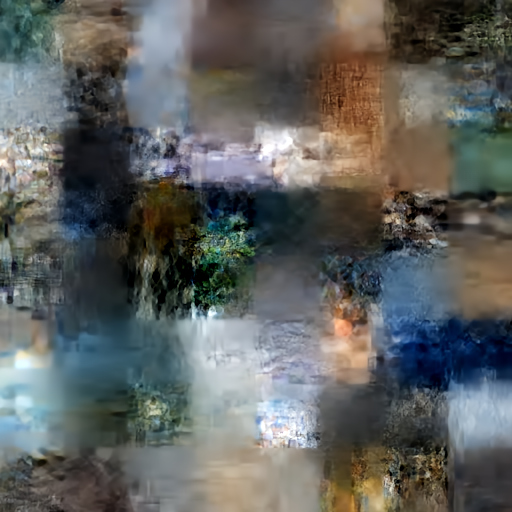}
    \vspace{\subcap}
    \caption*{Hyperprior\\(Mean \& Scale)~\cite{minnen2018neurips}}
  \end{subfigure}
  \hfill
  \begin{subfigure}[t]{\figwidth}
    \centering
    \includegraphics[width=\imagewidth]{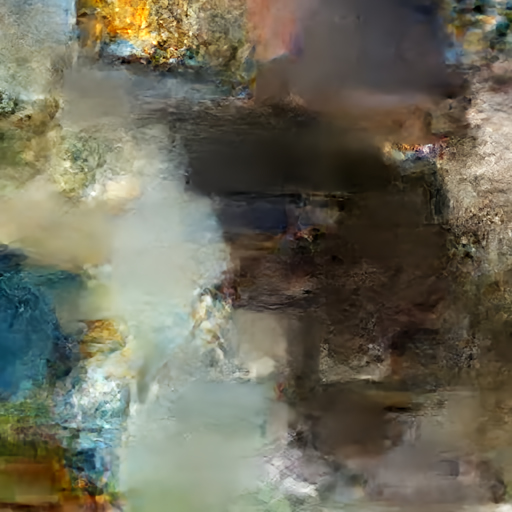}
    \vspace{\subcap}
    \caption*{Hyperprior +\\Spatial AR~\cite{minnen2018neurips}}
  \end{subfigure}
  \hfill
  \begin{subfigure}[t]{\figwidth}
    \centering
    \includegraphics[width=\imagewidth]{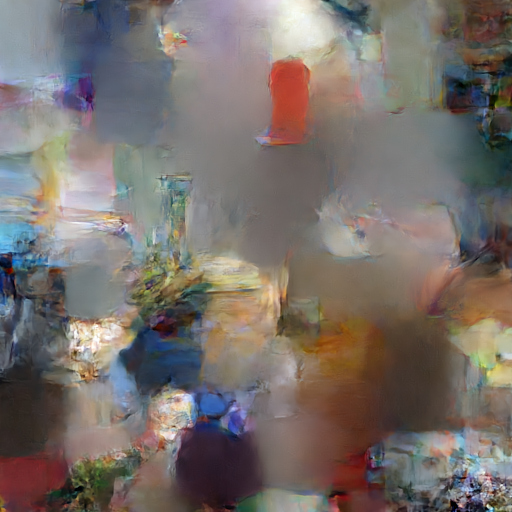}
    \vspace{\subcap}
    \caption*{Channel-wise AR Model (this paper)}
  \end{subfigure}
  
  \vspace{-5pt}
  \captionsetup{justification=justified}
  \caption{Random samples from different learned image compression models in order of increasing RD performance. All of the models are fully convolutional, and block artifacts due to strided convolution are clearly visible, especially in the earlier, less sophisticated models.}
  \label{fig:samples-different-models}
\end{figure*}

To better understand the sampled images in the context of learned image compression, Figure~\ref{fig:samples-different-models} shows typical samples drawn from six different architectures. All of the models are fully convolutional and use strided convolution to reduce the spatial extent of the data in deeper layers of the network. This downscaling creates block artifacts in the sampled images since all of the models use a factorized entropy model at the highest level (\ie the samples are spatially i.i.d. at this level). For example, the Spatially Local Histograms~\cite{minnen2018icip} approach uses 16x downscaling to form a latent representation and models each $13 \times 13$ tile in latent space using its own histogram. This structure is evident in the sampled image since we can see the large blocks (corresponding to the local histograms) and the smaller blocks within each tile (corresponding to 16x downscaling). Only the Spatial AR model~\cite{minnen2018neurips} is capable of long-range spatial dependencies, as is visible in the sample.

The channel-wise AR model is more limited than the spatially AR model in terms of long-range coherence. Nonetheless, the channel-conditional structure allows for a much larger receptive field than any of the earlier models since the receptive field can grow with each success slice of the latent representation.

All of the models visualized here were optimized for mean squared error (MSE). More research is needed to determine if the lack of any obvious semantic information in the samples is due to this simple loss function or if it's primarily due to insufficient capacity. Typically, generative models are optimized over constrained domains (faces, city views, flowers, bedrooms, \etc) and often use much larger networks than what is explored here. Interesting future work includes training our channel-wise AR model on a constrained domain and integrating an adversarial loss to see how each change affects the sampled images.

\begin{figure}[tb]
  \centering
  \includegraphics[width=\columnwidth]{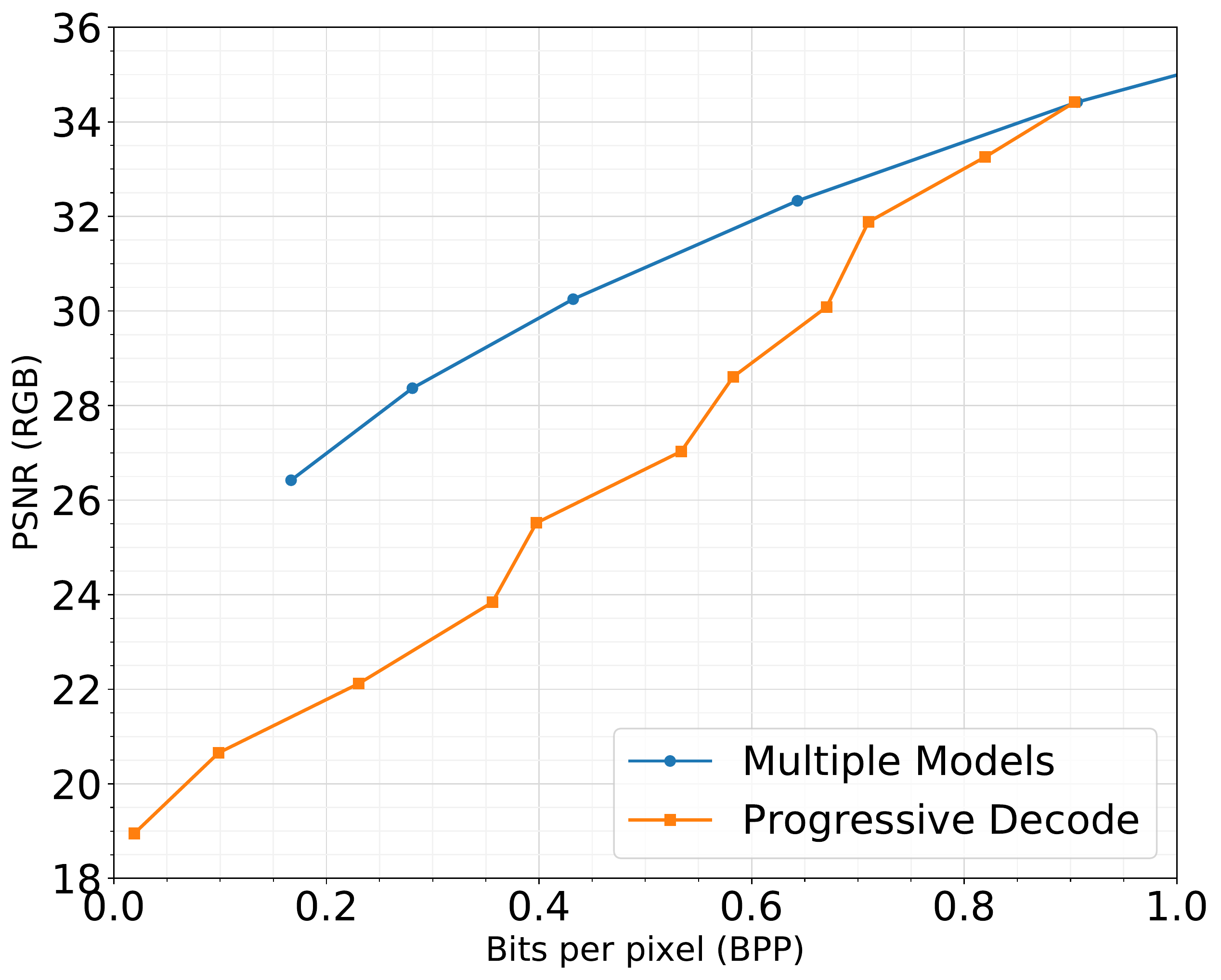}
  \shrinkcaption
  \caption{Our model supports progressive decoding by generating an image first based on just the hyperprior and then on each slice as it's decoded. While this may be useful for temporary previews, the rate-distortion performance is much worse than models separately optimized for lower bit rates.}
  \label{fig:progressive-rd}
\end{figure}

\begin{figure*}[tb]
  \setlength\figwidth{0.245\linewidth}
  \setlength\imagewidth{0.244\linewidth}
  \setlength\subcap{-20pt}
  
  \centering
  \begin{subfigure}[t]{\figwidth}
    \centering
    \includegraphics[width=\imagewidth]{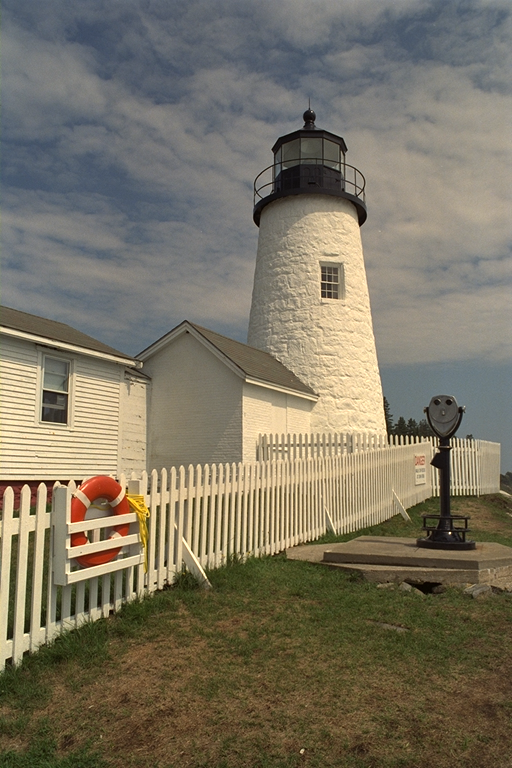}
    \vspace{\subcap}
    \caption{Original image}
  \end{subfigure}
  \hfill
  \begin{subfigure}[t]{0.496\linewidth}
    \centering
    \includegraphics[width=\imagewidth]{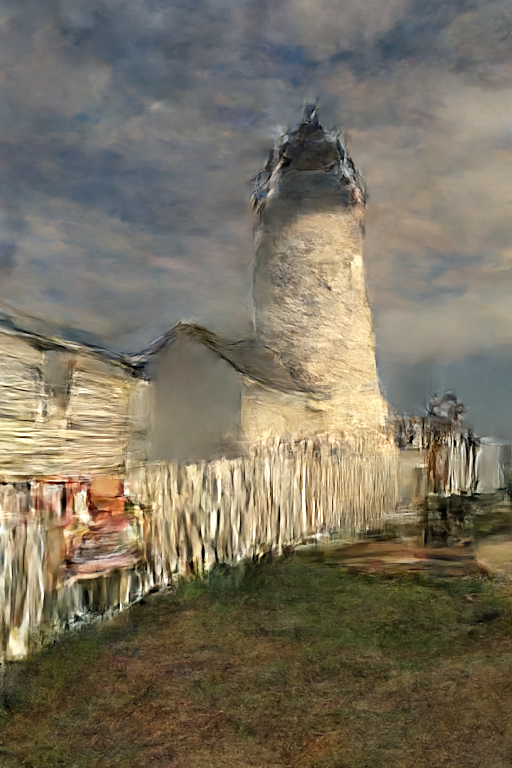}
    \hfill
    \includegraphics[width=\imagewidth]{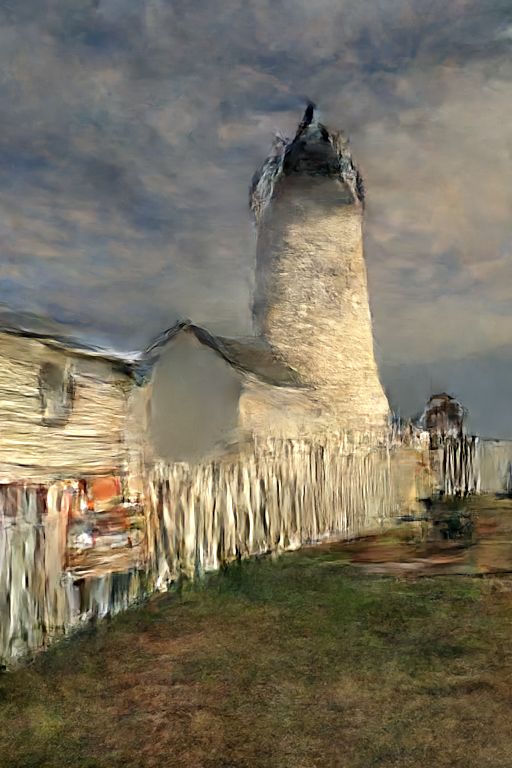}
    \vspace{\subcap}
    \caption{Two random samples from the entropy model}
  \end{subfigure}
  \hfill
  \begin{subfigure}[t]{\figwidth}
    \centering
    \includegraphics[width=\imagewidth]{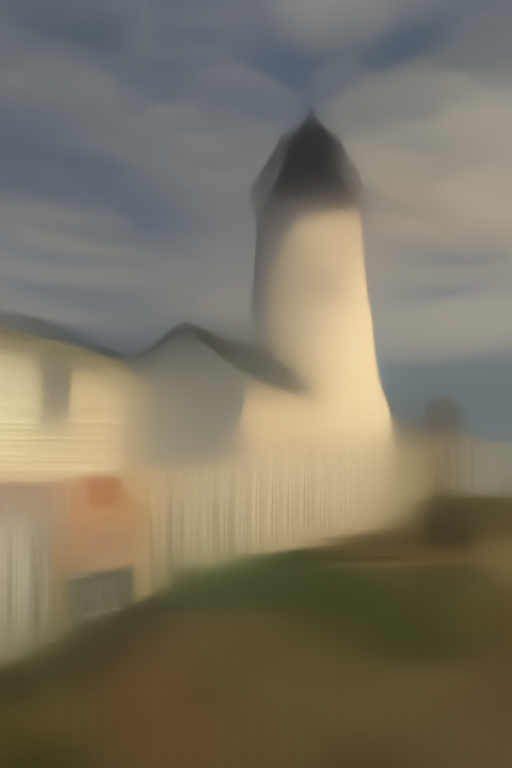}
    \vspace{\subcap}
    \caption{Mode of the entropy model}
  \end{subfigure}
  
  \vspace{-5pt}
  \caption{We can visualize the information stored in the hyperprior by using the hyperprior values from (a) a real image, and then (b) sampling from the conditional entropy model. Alternatively, we can (c) generate latents by taking the mode of the conditional entropy model rather than drawing random samples. The partial sampling shows that the hyperprior stores low frequency color data along with some texture and orientation information, \eg as demonstrated by the vertical fence pickets, horizontal slats on the building, and the different textures in the grass and sky.}
  \label{fig:samples-lighthouse-hyperprior}
\end{figure*}

\subsection{Samples with a Known Hyperprior}

To better understand what information is represented in the hyperprior, we can partially sample from the compression model. In this case, a real image is encoded and the resulting hyperprior is saved. We then repeat the sampling process described above but use this known hyperprior instead of a random one. Figure~\ref{fig:samples-lighthouse-hyperprior} shows such partial samples based on the lighthouse image (kodim19) from the Kodak image set~\cite{kodak}. In this example, the hyperprior requires 0.0143 bpp, which is an extremely compact representation (nearly 1680x compression) compared to typical rates used for image compression.

Two sampling approaches are explored. In the first (Figure~\ref{fig:samples-lighthouse-hyperprior}b), random samples are drawn from the entropy model conditioned on the real hyperprior. In the second approach (Figure~\ref{fig:samples-lighthouse-hyperprior}c), the mode of the conditional entropy model is used to form the latent tensor. Since the compression model uses a conditional Gaussian distribution and thus the mode is located at the mean, the difference between the methods is whether we sample from the Gaussian at each location or if we use the predicted mean.

From the images generated by this partial sampling procedure, we see that the hyperprior stores low frequency color data along with some texture and orientation information. For example, the strong vertical components in the fence are visible as well as the horizontal components in the slats on the building on the left side of the image. The hyperprior also stores a small amount of texture information as demonstrated by the different high-frequency patterns in the sky compared to the grass or the stone lighthouse.

\section{Progressive Decoding}

\begin{figure*}[tb]
  \centering
  \newlength{\rowspace}
  
  \setlength\figwidth{0.33\linewidth}
  \setlength\imagewidth{0.325\linewidth}
  \setlength\subcap{-10pt}
  \setlength\rowspace{4mm}
  
  \begin{subfigure}[t]{\figwidth}
    \centering
    \includegraphics[width=\imagewidth]{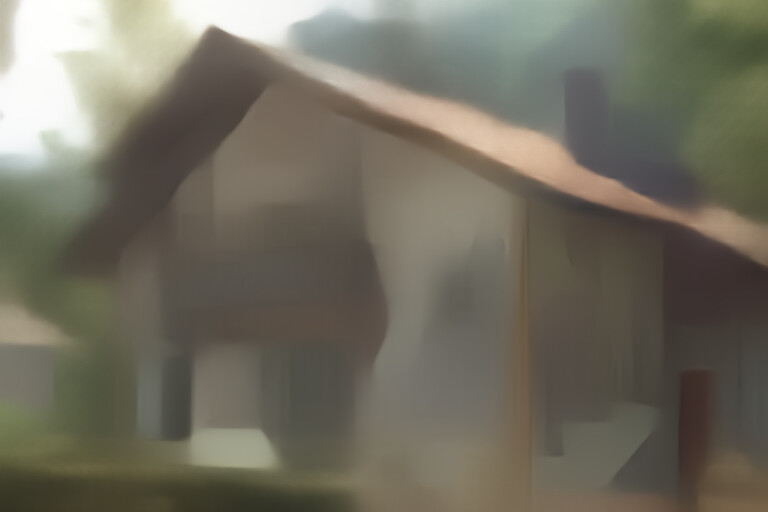}
    \vspace{\subcap}
    \caption{Hyperprior (0.01921 bpp, 18.95 dB)}
  \end{subfigure}
  \hfill
  \begin{subfigure}[t]{\figwidth}
    \centering
    \includegraphics[width=\imagewidth]{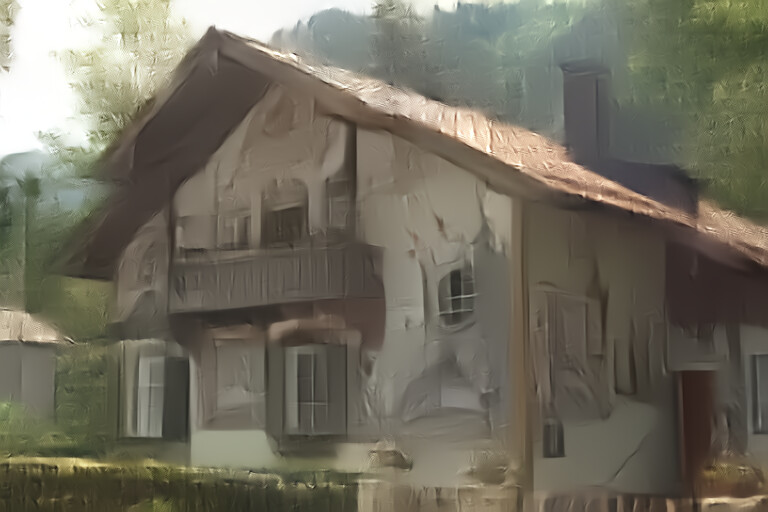}
    \vspace{\subcap}
    \caption{Slice 1 (0.09859 bpp, 20.66 dB)}
  \end{subfigure}
  \hfill
  \begin{subfigure}[t]{\figwidth}
    \centering
    \includegraphics[width=\imagewidth]{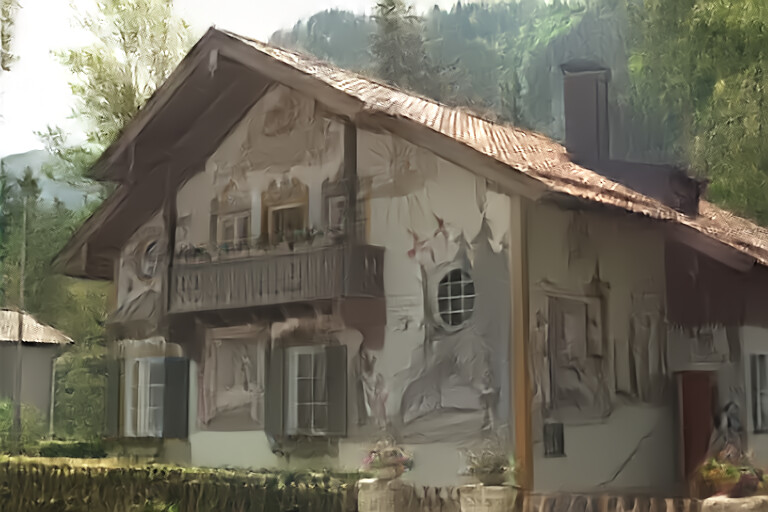}
    \vspace{\subcap}
    \caption{Slices 1--2 (0.23047 bpp, 22.12 dB)}
  \end{subfigure}
  \\
  \vspace{\rowspace}
  \begin{subfigure}[t]{\figwidth}
    \centering
    \includegraphics[width=\imagewidth]{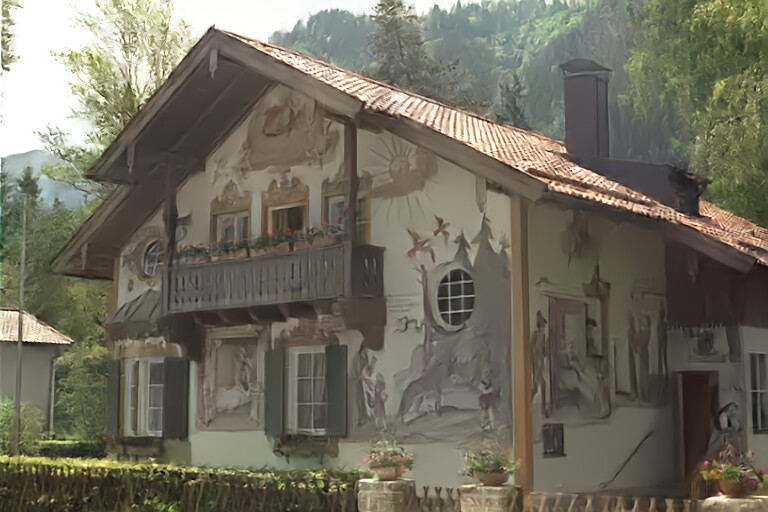}
    \vspace{\subcap}
    \caption{Slices 1--3 (0.35612 bpp, 23.84 dB)}
  \end{subfigure}
  \hfill
  \begin{subfigure}[t]{\figwidth}
    \centering
    \includegraphics[width=\imagewidth]{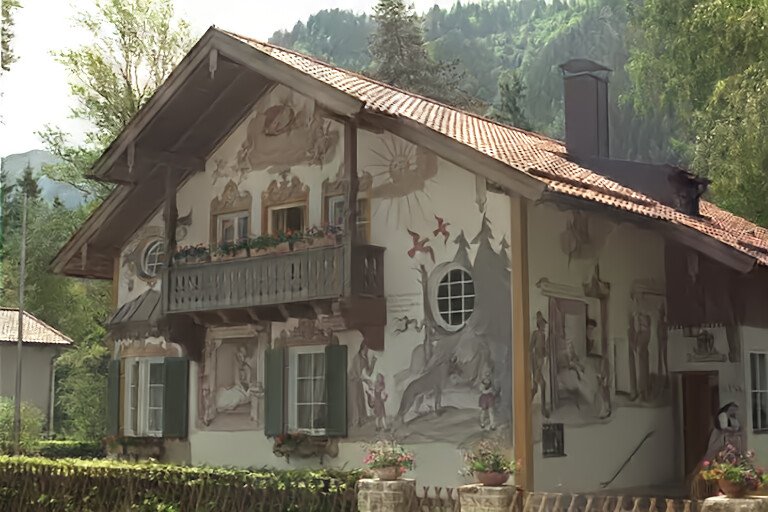}
    \vspace{\subcap}
    \caption{Slices 1--4 (0.39758 bpp, 25.52 dB)}
  \end{subfigure}
  \hfill
  \begin{subfigure}[t]{\figwidth}
    \centering
    \includegraphics[width=\imagewidth]{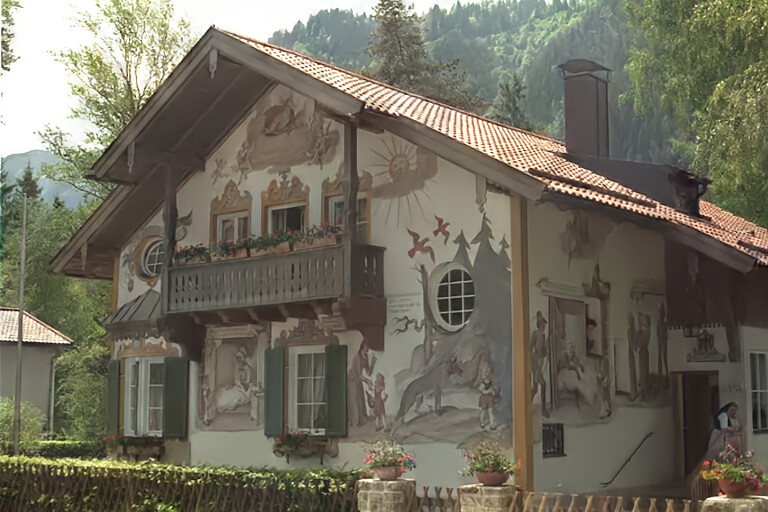}
    \vspace{\subcap}
    \caption{Slices 1--5 (0.53369 bpp, 27.03 dB)}
  \end{subfigure}
  \\
  \vspace{\rowspace}
  \begin{subfigure}[t]{\figwidth}
    \centering
    \includegraphics[width=\imagewidth]{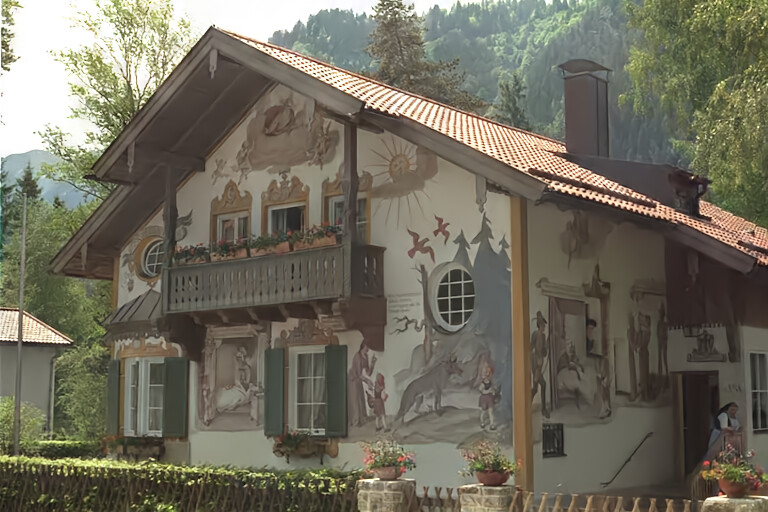}
    \vspace{\subcap}
    \caption{Slices 1--6 (0.58264 bpp, 28.61 dB)}
  \end{subfigure}
  \hfill
  \begin{subfigure}[t]{\figwidth}
    \centering
    \includegraphics[width=\imagewidth]{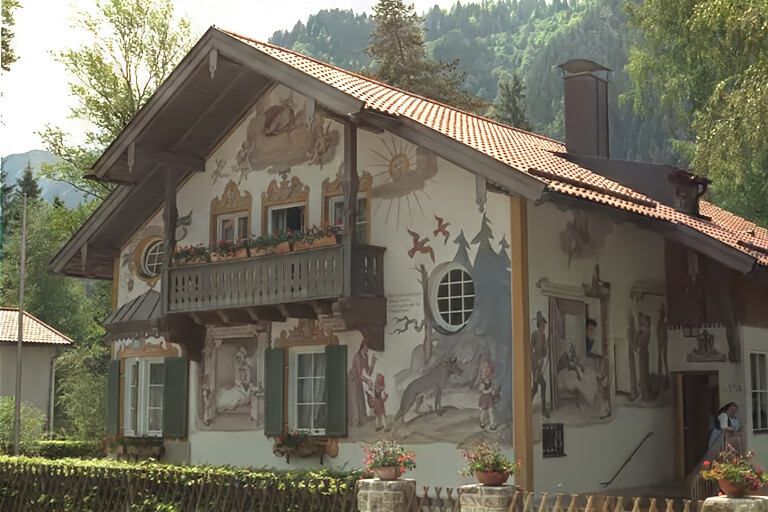}
    \vspace{\subcap}
    \caption{Slices 1--7 (0.67053 bpp, 30.08 dB)}
  \end{subfigure}
  \hfill
  \begin{subfigure}[t]{\figwidth}
    \centering
    \includegraphics[width=\imagewidth]{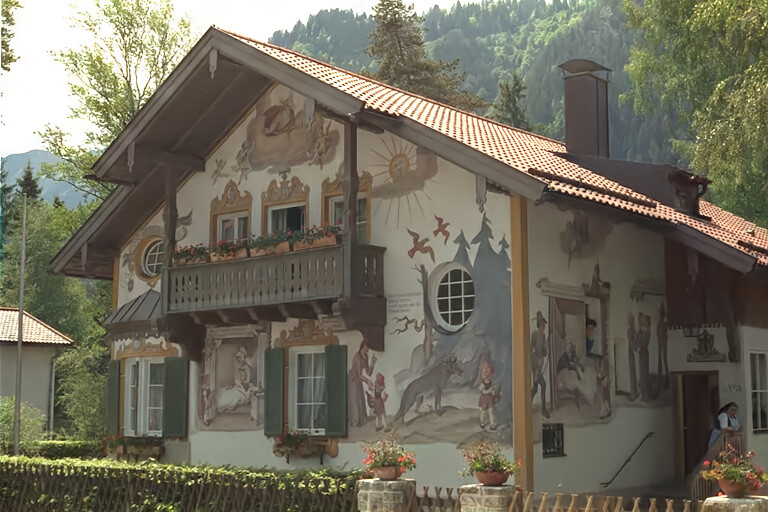}
    \vspace{\subcap}
    \caption{Slices 1--8 (0.70988 bpp, 31.88 dB)}
  \end{subfigure}
  \\
  \vspace{\rowspace}
  \begin{subfigure}[t]{\figwidth}
    \centering
    \includegraphics[width=\imagewidth]{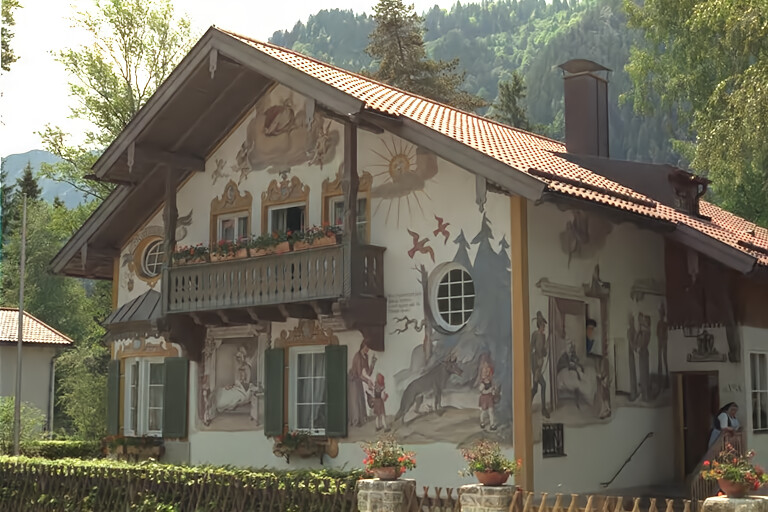}
    \vspace{\subcap}
    \caption{Slices 1--9 (0.81966 bpp, 33.26 dB)}
  \end{subfigure}
  \begin{subfigure}[t]{\figwidth}
    \centering
    \includegraphics[width=\imagewidth]{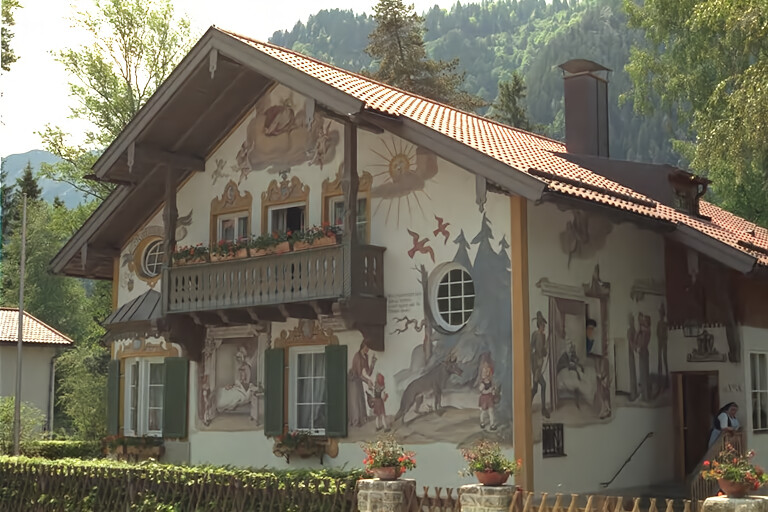}
    \vspace{\subcap}
    \caption{Slices 1--10 (0.90397 bpp, 34.42 dB)}
  \end{subfigure}

  \vspace{-1pt}
  \caption{For a 10-slice channel-conditional model, we can progressively decode the latent tensor to generate 11 images. The first image is based solely on the hyperprior. The next ten images are based on the hyperprior plus the first N slices of the latent representation. Missing values, \ie values for slice$_{i+1}$ when only slices 1--$i$ are available, are filled with the mode from the conditional distribution inferred from the hyperprior.}
  \label{fig:progressive-decode}
\end{figure*}

Although the primary purpose of our channel-wise autoregressive model is to improve entropy coding, the structure naturally lends itself to progressive decoding. A rough image can already be displayed after the hyperprior is transferred as shown in Figure~\ref{fig:samples-lighthouse-hyperprior}c. This reconstruction can then be improved after each slice in the latent space is decoded. Figure~\ref{fig:progressive-decode} shows the sequence of reconstructions recovered from a 10-slice model where the synthesis transform is executed after each slice is decoded, and missing latent values use the mode of the conditional distribution inferred from the hyperprior.

While progressive decoding is possible and easy to achieve using our channel-conditional model, two issues make it fairly impractical. First, the rate-distortion curve implied by the progressively decoded images is much worse than the result from separately optimized models (see Figure~\ref{fig:progressive-rd}). This means that progressive decoding may be useful as a temporary preview, but it does not lead to an effective multi-rate model. Second, the computational cost is fairly high since the full synthesis transform must be run to generate each image. A more useful model would reduce the computational cost along with the bit rate to support progressive previews.


\clearpage


\clearpage

\clearpage

\clearpage


\clearpage
\end{appendices}

\vfill
\clearpage

\renewcommand*{\bibfont}{\small}
\printbibliography

\end{document}